\documentclass[aps,prl,twocolumn, superscriptaddress]{revtex4-1}

\usepackage{color}
\usepackage{xcolor}
\usepackage{amsmath}
\usepackage{amssymb}
\usepackage{subfigure}

\usepackage{graphicx}

\newcommand\x{\times}

\begin{document}

%\title{Projected Zero LL in pairs of Weyl Semimetal with opposite chiralities}
%\title{Persistence of Zero Landau Energies Protected from Projected Vorticities of Opposite Chiral Weyl Nodes}
%\title{Persistence of Zero Landau Energies Protected from Vorticities of Opposite Chiral Weyl Nodes}
\title{
Topologically distinct Weyl-fermion pairs: on the effect of magnetic tunnelling}

\author{Ming-Chien Hsu}
%\email[E-mail address: ]{mingchienh@gmail.com}
\affiliation{Department of Physics, National Sun Yat-sen University, Kaohsiung 80424, Taiwan}	
										  
\author{Hsin Lin}
\affiliation{Institute of Physics, Academia Sinica, Nankang Taipei 11529, Taiwan}

\author{M.\ Zahid Hasan}				
\affiliation{Laboratory for Topological Quantum Matter and Spectroscopy (B7), 
Department of Physics, Princeton University, Princeton, New Jersey 08544, USA}
			
\author{Shin-Ming Huang}
\email[Corresponding author: ]{shinming@mail.nsysu.edu.tw}
\affiliation{Department of Physics, National Sun Yat-sen University, Kaohsiung 80424, Taiwan}

\begin{abstract}
A Weyl semimetal has Weyl nodes that always come in pairs with opposite chiralities. 
Notably, different ways of connection between nodes are possible and would lead to distinct topologies. Here we identify their differences in many respects from two proposed models with different vorticities. One prominent feature is the behaviour of zeroth Landau levels (LLs) under magnetic field. 
We demonstrate that the magnetic tunnelling does not always expel LLs from zero energy because the number of zero-energy modes is linked to the vorticity of the Weyl nodes, instead of the chirality. Other respects in disorder effects for weak (anti-)localization, surface Fermi arcs, and Weyl-node annihilation, are interesting consequences that await future exploration.
\end{abstract}

\maketitle

%\section{Introduction}
\emph{Introduction.}---
The prediction of Weyl semimetals (WSMs)
 \cite{WS11_FA, TImultiLay11, TRI_WS12, WS11_HgCr2Se4, WS14_DFT_LaBiTe, Turner2013} and its realization in real materials
 \cite{Huang2015, WS15_FA, TaAs15_zFang, TaAs15a_zFang, Yang2015, NbAs15a_ShinMing, WSexp_FA15} make the relativistic chiral fermions find their counterpart in three-dimensional (3D) condensed matter systems. These chiral fermions reside in the nodes of the electronic structure around which the energy dispersion is linear in momentum, the so-called Weyl nodes (WNs). The WNs always come in pairs with opposite chiralities \cite{Nielsen-Ninomiya81a, Nielsen1983adler} which act as the source and the sink of the Berry curvature \cite{ARev17_WS}. Due to topological reasons, various unusual behaviours were found in the WSM, such as the chiral anomaly \cite{Anamaly69, chAnomaly_WF83, Burkov12_chiralAnomaly, 
ChiralAno_Vishwanath14, Jia2016}, negative magnetoresistance \cite{NMR_th13, NMR15a_SdH}, chiral magnetic effect \cite{PRB14_CME}, novel quantum oscillation \cite{PRB12_FriedelOsci, Film_FA14_Vishwanath, Film_FA16_Vishwanath, PRB16.93.081103}, 
Fermi arc from the surface states \cite{WS11_FA, FA15_spinTexture, Jia2016, science16_STM_FA}, and so on.

One prominent and important phenomena of the WN is the behaviour of Landau levels (LLs) under magnetic field, which were mapped out by magneto-optical study recently in NbAs \cite{chiral_LL18}.
 The zeroth LL ($n=0$), also called chiral Landau band, has linear dispersion along the field direction, say $\hat{z}$, as $E=\chi v_z k_z$. Interestingly, the chirality $\chi =\pm 1$ determines the sign of the band slope. In particular, the connection of opposite chiral bands by field applied along the WNs provides the platform of chiral anomaly in which the charge pumping breaks the chiral symmetry. This actually relies on the existence of the zero-energy modes at $k_z=0$ which are topologically protected, or otherwise the system becomes insulating and the charge pumping may be forbidden.
 
Semiclassically the LLs are formed through quantization conditions of cyclotron orbits. When cyclotron orbits encounter each other, go across density discontinuities \cite{Tunnel_BC88}, or are blocked by some boundaries \cite{CorbinoTunnel13}, different quantization conditions may be formed and hence the LLs are changed. Besides, there exists tunnelling between cyclotron orbits, generally known as magnetic breakdown \cite{shoenberg_1984}. For example it has been discovered in adjecent quantum wells \cite{CorbinoTunnel13}. Particularly, inter-level tunnelling between levels from separate chiralities produce new features in transport in graphene \cite{Graphene_tunnel15}. Since level mixing is common, the question of interest now is whether the phenomenon happens in WSMs. As WNs are connected by bands, the magnetic tunnelling between LLs is expected when the field is applied perpendicular to the connection of WNs, possibly gapping the zeroth LL.
If it is the case, this would also lead to failure of chiral anomaly, as indicated as the possible explanation for the increased magneto-resistance \cite{1705.00920, QMlimit_McDonald18} or sharp sign reversal of Hall resistivity \cite{MagTunnel_TaP17}.
 
However, the reverse inference might not be true, meaning that the magneto-resistance changes are not necessarily attributed to the gapping of zeroth LLs.
They may also be caused by the gap opening in the system through
multiple Weyl carrier interaction \cite{QMlimit_McDonald18} or with the help of other non-Weyl singularities \cite{PhysRevLett.119.266401}. 
Besides, recently there are also concerns about interpreting the measured negative longitudinal magnetoresistance as direct evidence of chiral anomaly \cite{NJP17_misalignmentNMR, Li2017_extrinsicNMR, PRX.8.031002}.
Therefore, not only helping identify reasons of resistance changes affirmatively but also with theoretical significance, it is important and interesting to see if the gapping of zeroth LLs is an inevitable result.
Actually, the above inference of repelling zero modes only considers the simple connection between WNs, while symmetry constraints could make situations change. 
For example, the mirror symmetry is the commonly seen constraint connecting the WNs \cite{Huang2015, WS15_FA, TaAs15_zFang, WSexp_FA15, PRB.93.205109}.

To consider consequences from different ways of connection between WNs, we studied two models with different symmetry constraints imposed.
Magnetic tunnelling was found to be different and the zero-energy modes are still robust in one model against the tunnelling.
We attribute the findings to distinct topological invariants, suggesting that chirality alone is not sufficient to characterize a WN. 
Moreover, respective unique phenomena in impurity scattering, surface Fermi arcs, and WN annihilation are also studied and can differentiate the two topological distinct models.
The mutual annihilation between WNs does not necessarily become gapped and a nodal ring is possible to be formed, consistent with the newly discovered conversion rule \cite{PRL.121.106402}.
Therefore, the detailed ways of connection between WNs await future more investigation.

%\section{model description}
\emph{Models.}---
% ------------------------- %
%	model (II) description	%
%		w/o mirror symmetry	%
% ------------------------- %
We consider a pair of WNs with opposite chiralities sitting on two sides of the mirror plane $M_{x}$. Their separation $2k_\mathrm{W}$ is relatively small compared to the size of the Brillouin zone (BZ) such that the magnetic field has a chance to couple them. Other WNs, if exist, in other regions of the BZ can be ignored as they are much far away. 
To have 
\begin{equation}
\mathcal{M}_{x} H(k_x,k_y,k_z)\mathcal{M}_{x}^{-1} = H(-k_x,k_y,k_z),
\end{equation}
we find that there are two possible choice of $\mathcal{M}_{x}$ and then models. We dub them Model A and Model B: Model A is for $\mathcal{M}_{x} = \sigma_{0}$ that the two bands have equal mirror parity in the mirror plane, while Model B is for $\mathcal{M}_{x} = \sigma_{x}$ that two bands have opposite mirror parities. Specifically, the Hamiltonian are written as
\begin{equation}
H_{\mathrm{A}} = \frac{1}{2m}\left(k_x^2 - k_\mathrm{W}^2 - \alpha k_\parallel^2 \right)\sigma_x + v_{\parallel} k_y \sigma_y 
      + v_{\parallel} k_z \sigma_z, 
      \label{model_H_A}
\end{equation}
\begin{equation}
H_{\mathrm{B}} = \frac{1}{2m}\left(k_x^2 - k_\mathrm{W}^2 - \alpha k_{\parallel}^2 \right)\sigma_x 
 		 + \frac{v_\parallel}{ k_\mathrm{W}}  k_x \left( k_y \sigma_y + k_z \sigma_z \right),
 		\label{model_H_B}      
\end{equation}
where $k_\parallel^2 = k_y^2 + k_z^2$.
In general, coefficient for $k_y$ and $k_z$ can be different, but the physics are the same. The Planck constant $\hbar$ is set as unity throughout the paper.
We note that the origin is meaninglessly specified and might not be at a time-reversal-invariant momentum. 

Both two models contain two WNs located at $(\pm k_\mathrm{W}, 0, 0)$. Expanding around the WNs, they approximate, to linear order, as
\begin{eqnarray}
H_{\mathrm{A}} &\approx & \sum_{\chi = \pm} \left( \chi v_{x} q_{x}\sigma_x + v_{y} q_{y} \sigma_y  + v_{z} q_{z} \sigma_z \right), 
      \label{model_H_A1} \\
H_{\mathrm{B}} &\approx & \sum_{\chi = \pm} \chi \left(  v_{x} q_{x}\sigma_x + v_{y} q_{y} \sigma_y  + v_{z} q_{z} \sigma_z \right). 
 		\label{model_H_B1}      
\end{eqnarray}
Here $\chi$ labels the chirality and also position of the WN, and $v_{x}=k_\mathrm{W}/m$ and $v_y = v_z = v_\parallel$.
The values of $k_\mathrm{W}$, $m$, $v_y$ and $v_z$ are all assumed to be positive without loss of generosity. $\alpha$ ($>0$) in $H_{\mathrm{B}} $ is required in order to have a saddle point at energy $E_{\mathrm{VH}}=k_{\mathrm{W}}^2/2m$, above which close energy contours are assured. Although a $k_\parallel$-linear term is allowed to appear in the $\sigma_{x}$ term, it is omitted for an elegance reason. We have confirmed that its presence once being small does not change qualitative conclusions. 

Without special regard to the symmetry or phase, Model A was usually adopted to study the effect of a pair of nodes  \cite{MagTunnel_TaP17,Chan2017a}. 
Applying the magnetic field along the perpendicular $z$ direction to this system, we solve the LL spectrum by substituting $\mathbf{k}$ in Eq. (\ref{model_H_A}) by $\mathbf{\Pi} = \mathbf{k} + e \mathbf{A}$. We choose the Landau gauge $\mathbf{A}= Bx\hat{y}$, so we make $k_x\rightarrow \Pi_x=k_x$, $k_z\rightarrow \Pi_z=k_z$, and 
$k_y \rightarrow \Pi_y = l_B^{-2}\bar{x}$,
where the magnetic length $l_B = \sqrt{\frac{1}{eB}}$ and $\bar{x}$ is the coordinate relative to the guiding center $x_0= - l_B^2 k_y$. $\bar{x}$ is conjugate to $k_x$ by quantizing $\bar{x} \rightarrow i\frac{\partial}{\partial k_x}$.
The WN separation is used as a measure to define the dimensionless momentum scale as $q = k_x/k_\mathrm{W}$. 
Since the magnetic field breaks the inplane translation symmetry, two WNs are expected to couple via the field. The coupling way can be revealed from the missing terms to Eq. (\ref{model_H_A1}). We use a dimensionless parameter $g$ to describe the degree of the coupling. The coupling will increase with the cyclotron energy $\omega_{c}=\sqrt{2 v_x v_y} l_{B}^{-1}$ and decrease with the energy barrier $E_{\mathrm{VH}}$, defined as $g= \frac{\omega_{c}^2}{4E_{\mathrm{VH}}^2}$.
$g$ is proportional to the magnetic field $B$,
and the appreciable coupling $g\approx 1$ is achieved when the magnetic length $l_B$ is comparable to the scale defined by $k_\mathrm{W}^{-1}$  %\cite{lengthscale}.
\footnote{Take Weyl semimetal TaAs for example. The nodes separation of $W_1$ is $k_\mathrm{W}=0.0072\left(\frac{2\pi}{a}\right)$, and the Fermi velocity in the conduction band is $(v_x, v_y, v_z) = (2.5, 1.2, 0.2)\times 10^5$  m/s \cite{PRB15.WSparm}. The coupling $g=1$ under the field strength $B\approx 11.88$ T having magnetic length $l_B=7.44$ nm, which is close to the length 7.60 nm defined by $k_\mathrm{W}^{-1}$.}
By defining the remaining variables into dimensionless quantities, 
$q_z = v_{\parallel} k_z \left( \frac{\omega_c^2}{4 E_\mathrm{VH}}\right)^{-1} =  (\frac{2}{g})(\frac{v_\parallel}{v_x})(\frac{k_z}{k_\mathrm{W}})$, and $\tilde{\alpha}=\left( \frac{v_x}{v_\parallel} \right)^2 \alpha $, we can study the Hamiltonian under magnetic field as a function of $q$ and $q_z$.
We numerically solve this system with raising and lowering operators. Special treatment is developed to solve it more efficiently and the details are shown in the Supplemental Material \cite{SuppM}.

% ----------------------------- %
%  evolution with respect to g    %
% ----------------------------- %
% ------------- %
%	figures		     %
% ------------- %
\begin{figure}[tb]
\begin{center}
\includegraphics[width=0.5\textwidth]{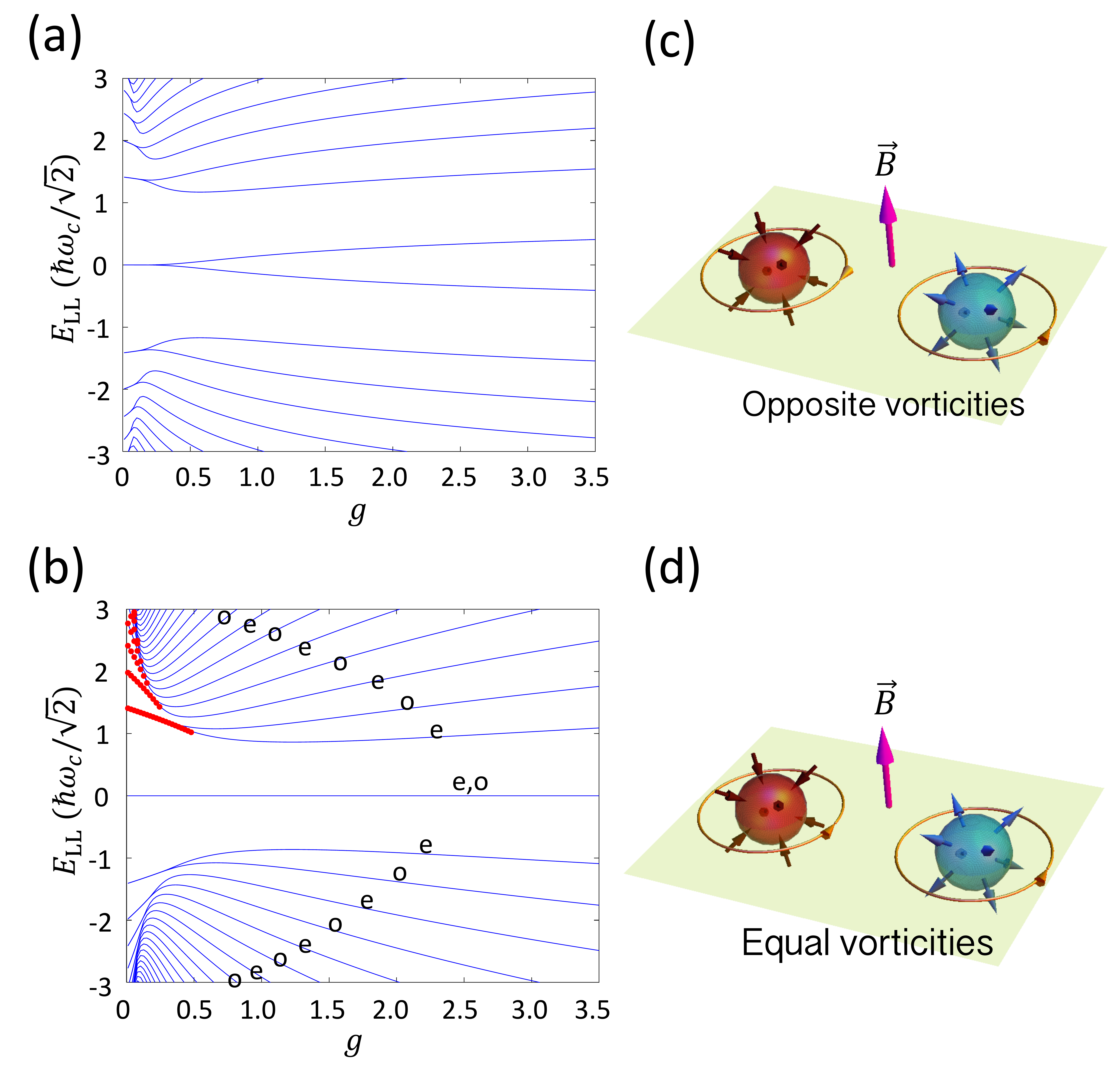}	
\caption{The LL spectra at $k_z=0$ with respect to the coupling measure $g$ with $\tilde{\alpha}=0.05$ for both Model A (a), and for Model B (b). 
In (b), Labels "e" and "o" denote even and odd parity of states, and the red dots are the analytical solutions, except zero-energy ones, when $\alpha=0$. (c,d) Illustrations of topological numbers of Weyl nodes, including chirality and vorticity, for Model A (c) and Model B (d). Chirality is identified by the outward or inward Berry flux, while vorticity is the directional winding number around a Weyl node.
\label{fig:EL_g}}
\end{center}
\end{figure}

The LL spectrum at $k_z=0$, i.e. $q_z=0$, with respect to $g$ is shown in Fig.~\ref{fig:EL_g}(a). In the limit of $g\rightarrow 0$, two WNs have independent and identical LLs, so each LL is doubly degenerate. As $g$ is turned on, the degeneracy is lifted off and band splits are visible at $g \approx 0.3$ (decrease with levels) in Fig.~\ref{fig:EL_g}(a). We have ascribed the band splits to the magnetic tunnelling in Ref. \cite{MagTunnel_TaP17} as the cyclotron wave functions in $k$ space broaden with the $B$ field and hybridize with others when overlaps occur.
Notice that the chiral symmetry is present for $\left\lbrace H,\sigma_z \right\rbrace=0$ at $k_z=0$, so the spectrum is symmetric about zero energy. In order to obey the chiral symmetry, the zeroth LL has to split into one with positive energy and one with negative energy. 

% --------------------------------- %
%   model (I)  w/ mirror symmetry	%
% --------------------------------- %
The results shown in Model A are reasonable but not conclusive. We solve Model B as follows. In Model B, the Peierls substitution should be carefully treated in $k_x k_y$ and $k_x k_z$. To make the Hamiltonian Hermitian, we do the symmetrization 
\[ k_x k_y \rightarrow \frac{1}{2} \left( \Pi_x \Pi_y + \Pi_y \Pi_x \right) 
	= l_B^{-2}\left( k_x\bar{x}+\frac{i}{2} \right),
\]
and $k_x k_z \rightarrow (\Pi_x \Pi_z + \Pi_z \Pi_x)/2= k_x k_z$. 
In terms of dimesionless parameters defined above,
the Hamiltonian under magnetic field becomes
\begin{equation}
 H_{\mathrm{B}}^{\prime} = \frac{\omega_{c}^2}{4E_{\mathrm{VH}}} \left\{
	 \begin{array}{l}
	 \left[\frac{1}{g}(q^2-1) 
	  + g\frac{\tilde{\alpha}}{4}\frac{\partial^2}{\partial q^2}
	  -  g\frac{\tilde{\alpha}}{4} q_z^2         \right] \sigma_x		\\
       + i \left(q\frac{\partial}{\partial q} + \frac{1}{2} \right)\sigma_y 
   	  + q_z q\sigma_z 
   	 \end{array}	 \right\},
\label{Eq:B1}
\end{equation}
where the prime stands for a system under a magnetic field.

The LL spectrum with respect to $g$ for Model B at $k_z=0$ is shown in Fig.~\ref{fig:EL_g} (b). For small $g$, the effect of $\tilde{\alpha}$ is small, so we provide the results for $\tilde{\alpha}=0$ as shown by red dots in Fig.~\ref{fig:EL_g} (b). The solutions are analytical to be $E_n= \frac{\omega_{c}^2}{2E_{\mathrm{VH}}} \sqrt{n \left(\frac{1}{g}-n \right) }$ with $n \in \mathbb{Z}_\ge 0 $ and each energy has twofold degeneracy. The LL energies emerge from the typical LL spectrum as $E_n = \sqrt{n} \omega_{c}$ in the limit of $g \rightarrow 0$ and deform with finite $g$. However, the analytical solutions for $\tilde{\alpha}=0$ are only applicable in a low-energy region. At high energies, with large $g$ or $n$, the LL quantization makes no sense when $E>E_{\mathrm{VH}}$ due to the flat dispersion along the $k_y$ axis and thus unbounded equal-energy contours. Specifically, the analytical solutions are applicable when
$n \leq \frac{1}{2}\left(\frac{1}{g} - \frac{1}{2} \right)$.

Since the model is invariant under inversion in $q$, the eigenstates will be either even (e) or odd (o) in $q$ as denoted in Fig.~\ref{fig:EL_g} (b). The even- and odd states appear alternatively in energy, showing that they evolve from a degenerate spectrum for small $g$. We were unable to prove whether the degeneracy for $n>0$ is exact at small and finite $g$ but found it seeming to be in the applicable region of $\tilde{\alpha}=0$. The band splitting is reasonable as seen in a symmetric double-well with finite tunnelling probability where even and odd states have different energies.
By contrast, Model A does not have this symmetry and therefore its eigenstates do not respect this symmetry in $q$. 

Nevertheless, two zero-energy LL states persist for all values of $g$ and $\tilde{\alpha}$. They can be directly verified by solving the zero-eigenvalue problem. The eigenfunctions are found to be $\left(0, e^{-\kappa q^2} \Phi(q) \right)^{T} $, where $\Phi (q)$ can be either ${_1F_1}(-\frac{1}{4}\lambda;\frac{1}{2}; \xi^2 q^2)$ or $H_{\lambda/2}(\xi q)$, the former being the Kummer confluent hypergeometric function and the latter the Hermite polynomial, indicating a double degeneracy. Here $\kappa=\frac{1+\sqrt{1-\tilde{\alpha}}}{g\tilde{\alpha}}$, $\xi=\sqrt{\frac{2\sqrt{1-\tilde{\alpha}}}{g\tilde{\alpha}}}$ and $\lambda = \frac{2+g\sqrt{1-\tilde{\alpha}}}{-g\sqrt{1-\tilde{\alpha}}}$.
The derivation of the analytical form can be found in the Supplemental Material \cite{SuppM}.

% ---------------------------------------------- %
%   discuss the robutsness of the E=0 solution 	  %
%																	       %
%			vorticity of the projected plane	 			  %
% ---------------------------------------------- %
The zero-energy LL is topologically guaranteed once the topological charge is nonzero. Therefore it was regarded as legitimate that the zeroth LL gaps for two WNs of zero net chiarlity under a strong magnetic field, as what we see in Model A, Fig.~\ref{fig:EL_g} (a). The interpretation has to be corrected when the persistent zero-energy LLs is demonstrated in Model B which retains zero net chirality as Model A. 
Therefore, chirality will not be responsible for the zero-energy LLs. Still, the zero-energy modes should be dictated by topology. One can understand that the chirality is a high dimensional topological invariant and hence is not suitable for explaining the LL system, since the systems for $k_z=0$ is restricted to a 2D system perpendicular to the magnetic field. In a 2D manifold pierced by holes (Weyl nodes), a 1D topological number is the end. Because of the presence of chiral symmetry, the systems belong to the AIII symmetry class and are classified by the winding number, a $\mathbb{Z}$-type topological invariant~\cite{RMP_Ryu16}. In the paper, we dub it vorticity.
In the chiral basis, the  phase $\phi$ in the off-diagonal entry of the Hamiltonian characterizes the vorticity defined to be
$\nu=\frac{1}{2\pi}\oint_{S^1} d\mathbf{k}_\parallel\cdot\nabla \phi$, where $S^{1}$ is a loop enclosing a (or multiple) WN(s) on the $k_z=0$ plane. Referring to Eqs. (\ref{model_H_A1}) and (\ref{model_H_B1}), two WNs in Model A take opposite vorticities, but equal vorticity in Model B. (The sign of vorticity might change by changing the basis, but the relative sign between two is invariant.) We illustrate chiralities and vorticities of WNs for the two models as the conclusion in Figs.\ \ref{fig:EL_g} (c) and (d). Therefore, the net vorticity in Model A is $\nu_{\mathrm{A}}=0$ and is $\nu_{\mathrm{B}}=2$ in Model B. According to the index theorem \cite{index_PLB77, PhysRevLett.51.2077, PhysRevD.29.2375, PhysRevD.30.809} and generalizing it, the absolute value of vorticity can be witnessed by the number of zero-energy LLs in a magnetic field.
We have also examined a tilted field in the $y$-$z$ plane to strengthen this proof \cite{SuppM}.

\begin{figure}[tb]
\begin{center}
	\includegraphics[width=0.45\textwidth]{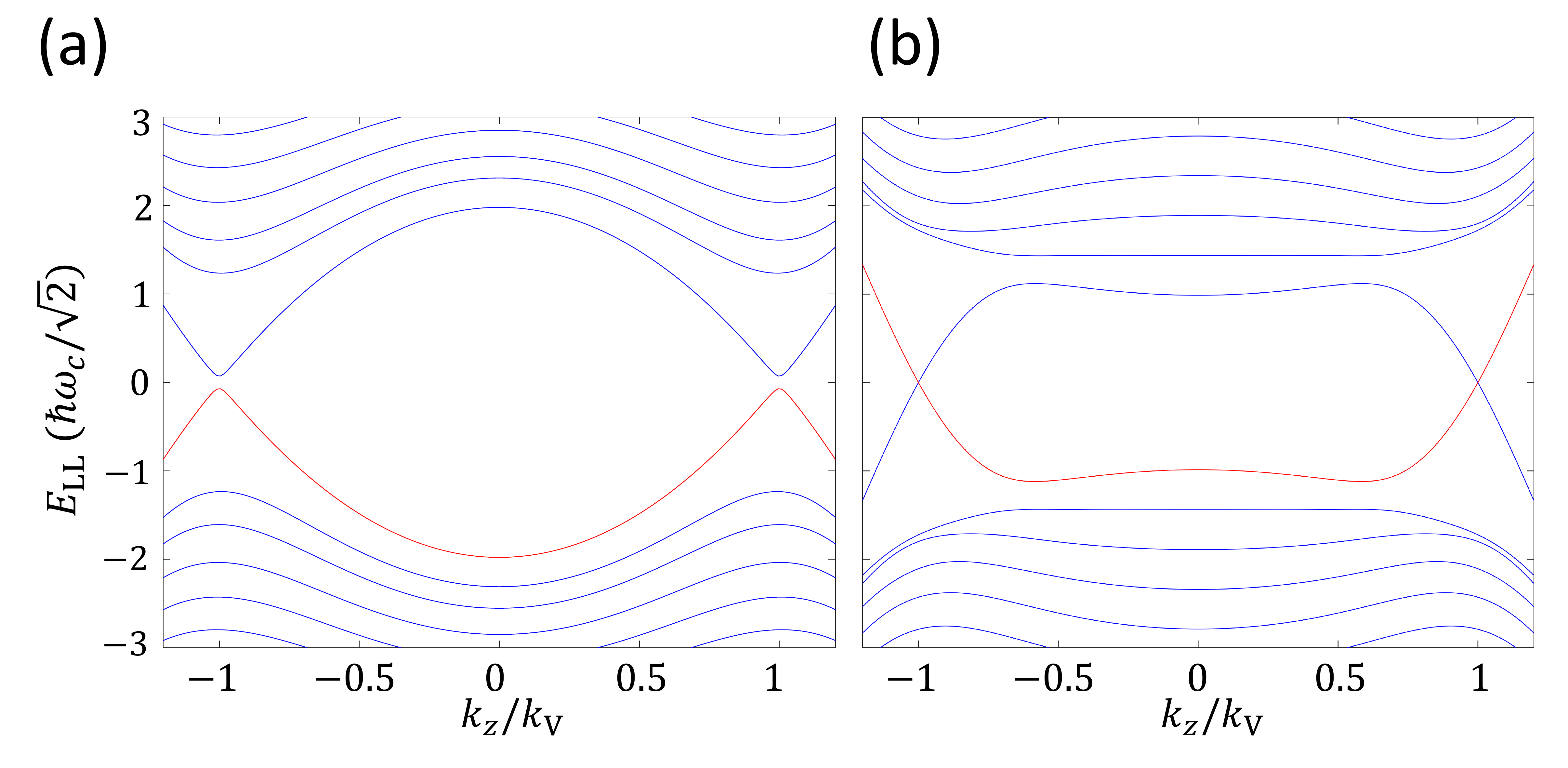}
\caption{The Laudau band spectrum with respect $k_z/k_{\mathrm{V}}$ for two pairs of Weyl nodes separated in the $k_z$ direction at $k_z=\pm k_{\mathrm{V}}$, respectively. (a) for Model A and (b) for Model B. In both (a) and (b), $g=0.8$, $\alpha=0.05$, $v_\parallel/v_x = 0.5$, and $\frac{k_{\mathrm{V}}}{k_\mathrm{W}}=0.2$ so that
$A_z =  (\frac{v_\parallel}{v_x})(\frac{k_{\mathrm{V}}}{k_\mathrm{W}}) = 2.5$.
\label{fig:Ekz_WS+-kz}
}
\end{center}
\end{figure}

%\section{Dispersion along $k_z$}
\emph{Dispersion along $k_z$.}---
% ------------------------------- %
%  evolution with respect to kz   %
% ------------------------------- %
The chiral anomaly is a phenomenon that in parallel magnetic and electric fields electric charges are transmitted from one WN to the other. To investigate this effect, we consider two pairs of WNs separated in $k_z$ with each pair as studied before. We modify our models by 
$k_z\rightarrow \frac{1}{2 k_{\mathrm{V}}}(k_z^2 - k_{\mathrm{V}}^2)$, 
and dub the modified models as Model $\tilde{\mathrm{A}}$ and $\tilde{\mathrm{B}}$ respectively; in this way the separation is $2k_{\mathrm{V}}$ and the absolute velocity component in $z$ is still $v_z$. 

We show the calculation results for Landau bands along $k_z$ in Fig.~\ref{fig:Ekz_WS+-kz}.
Since the magnetic field is along $z$, two pairs of WNs, at different $k_z$ connected in dispersion, are independent. In Model $\tilde{\mathrm{A}}$, as the zeroth LLs at both $k_z= \pm k_{\mathrm{V}}$ are gapped in large magnetic fields, a 3D insulating phase is present, as shown in Fig.~\ref{fig:Ekz_WS+-kz}(a). In Model $\tilde{\mathrm{B}}$, by contrast, the protected zero-energy states extend into the chiral Landau bands and result in a 3D metallic phase, Fig.~\ref{fig:Ekz_WS+-kz}(b). Moreover, the fact that the two chiral Landau bands crossing at either $k_z= k_{\mathrm{V}}$ or $-k_{\mathrm{V}}$ have opposite slopes is the proof of two WNs taking opposite chiralities. This feature reveals that to characterize the zeroth Landau bands in a WSM unequivocally, two topological invariants, chirality and vorticity, are necessary.

%\section{Impurity scattering}
\emph{Impurity scattering.}---
We then discuss other phenomena that may distinguish the two models which are characterized by the same chiralities but different vorticities. 
In weak magnetic fields, the conductivity is highly influenced by disorders. Contrary to normal materials, the topological semimetals undergo weak anti-localization in the absence of magnetic field due to the $\pi$-Berry phase from the WN to suppress backscattering \cite{Shen15_MR_WS}. The anti-localization phenomenon will fade away when inter-valley scattering is taken into consideration for the lack of topological protection. In chiral anomaly, the latter determines the scale of transport time. Therefore, inter-valley scattering will influence transport properties the most. 

We emphasize that two models have inherent distinction in inter-valley scatterings. 
Set $|v_x| = |v_y| = |v_z| \equiv v$ in Eqs. (\ref{model_H_A1}) and (\ref{model_H_B1}) for simplicity. Since each WN looks similar itself, the intra-valley scattering makes no difference between the two models. But for inter-valley scattering, whether the Fermi velocity changes sign or not from one valley to the other will affect the scattering probability. We denote the inter-valley scattering potential by $U^{+,-}_{q,q'}$ for a scattering from $q$ to $q'$ (momentum relative to WNs). Under Born approximation, the average scattering rate is given by 
\begin{equation}
\langle  \frac{1}{\tau_I} \rangle = \frac{2\pi}{\hbar}\sum_{q,q'}\langle |U^{+-}_{qq'}|^2\rangle \delta(E_F - \xi_{q'})\delta(E_F - \xi_{q})
\end{equation}
where $\xi_q=\hbar v_F q$. We realize that when the impurity is anisotropic as $p$-wave, the differences in two models will be identified. Take $p_y$-wave impurity for instance that the scattering potential $U^{+-}_{k,k'}\sim (q_y - q'_y)$ changes sign in $y$. As the Fermi velocity $v_y$ have opposite signs at two valleys in Model B, which indicates a $\pi$-phase difference between electrons at two valleys, inter-valley scattering will be enhanced by a $p_y$-wave impurity. In contrast, inter-valley scattering is weaker in Model A owing to equal sign of $v_y$. We conclude the results in Table~\ref{Tab:imp_tau}. 

\begin{table}
\begin{tabular}{|c|c|c|}
\hline
impurity potential  &  Model A                    & Model B \\
$U^{+-}_{q,q'}$        & $(+,+,+)\rightarrow(-,+,+)$ & $(+,+,+)\rightarrow(-,-,-)$  \\  \hline
  $u_I \frac{(q_x-q'_x)}{k_F}$    
  & $\langle \frac{1}{\tau_I} \rangle = \frac{2\pi}{\hbar} N_\mathrm{F} \frac{8}{9} u_I^2$ 
  & $\langle \frac{1}{\tau_I} \rangle = \frac{2\pi}{\hbar} N_\mathrm{F} \frac{8}{9} u_I^2$ 	\\	\hline
  $u_I \frac{(q_{\parallel}-q'_{\parallel})}{k_F}$    
  & $\langle \frac{1}{\tau_I} \rangle = \frac{2\pi}{\hbar} N_\mathrm{F} \frac{4}{9} u_I^2$ 
  & $\langle \frac{1}{\tau_I} \rangle = \frac{2\pi}{\hbar} N_\mathrm{F} \frac{8}{9} u_I^2$ 	\\	\hline   
\end{tabular}
\caption{Average inter-valley scattering rates $\langle \frac{1}{\tau_I} \rangle$ for anisotropic impurities potentials. Here $p$-wave ($p_x$, $p_y$, and $p_z$) impurities potentials are considered. 
The $p_x$-impurity potential (second row) does not differentiate the two models, but the form of $\sim p_\parallel=p_y$ or $p_z$ can tell the difference (last row).
$N_F$ is the density of states at the Fermi energy and $u_{I}$ characterizes the strength of the impurity potential.
\label{Tab:imp_tau}
}
\end{table}

\begin{figure}[tb]
\begin{center}
    \includegraphics[width=0.48\textwidth]{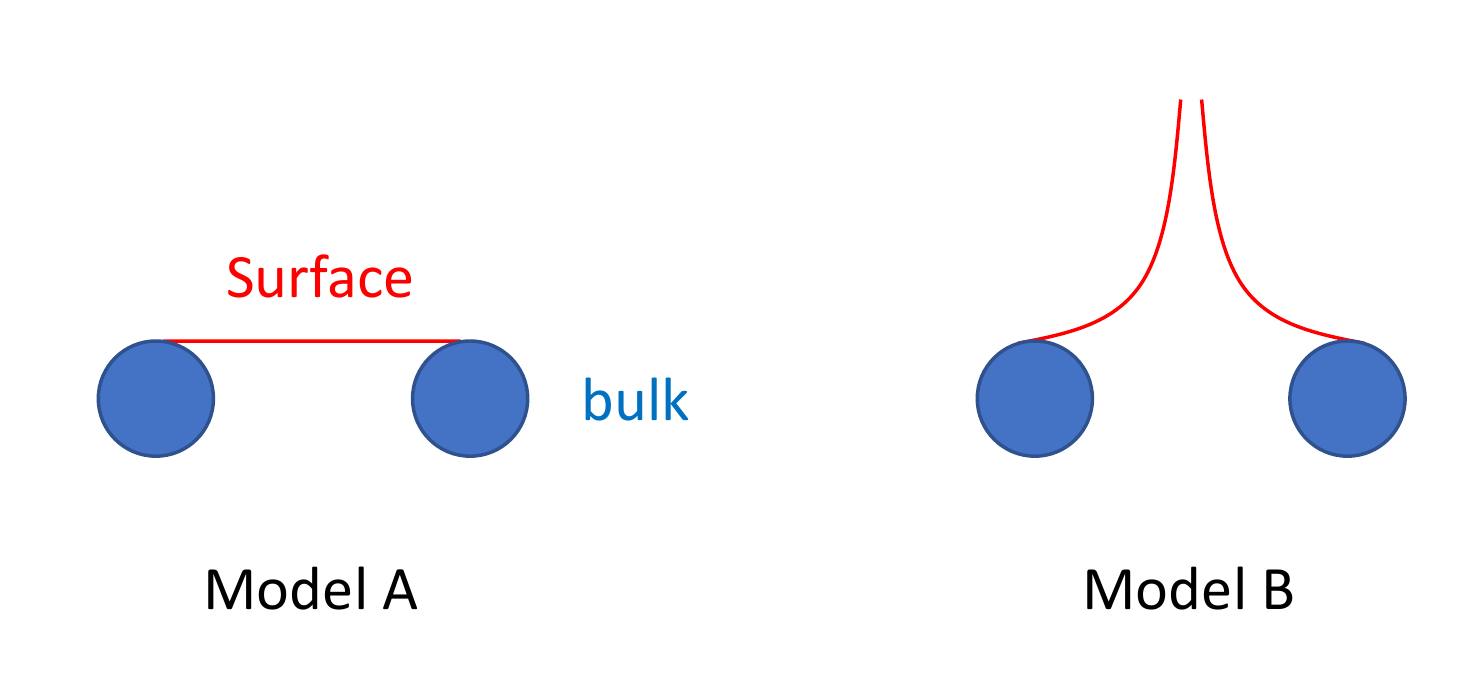}
\caption{The energy contour of surface states for Model A (left panel) and Model B (right panel). The blue areas stand for bulk projections (enclosing WNs) and red lines are for surface states. That Fermi arcs in Model B do not connect is visible.
\label{fig:surface}
}
\end{center}
\end{figure}

\emph{Surface states.}---
By solving WSM slabs with semi-infinity in the $z$ direction and a hard wall potential for $z>0$, we can have the surface states and the corresponding energies as a function of $(k_x, k_y)$. For Model A, the energy $E= -v_\parallel k_y$, and for Model B the energy is
 $E=\frac{v_\parallel}{k_\mathrm{W}}|k_x| k_y$. Therefore, the energy contours can be shown in Fig. \ref{fig:surface}, from which we found the Fermi arcs do not connect to each other in Model B, since the solved surface state wavefunction is not continous in $k_x=0$. (See details in Ref. \cite{SuppM}).
 
\emph{Weyl-node annihilation.}---
The two models differ in mirror parities of the two bands, so they give different results after the pair WNs collide and annihilate [by tuning $k_\mathrm{W}^2$ in Eqs. (\ref{model_H_A}) and (\ref{model_H_B}) to negative values].  After collision, WNs in Model A will gap the system while they evolve into a gapless nodal ring on the mirror plane in Model B \cite{SuppM}.
These are simply consequences of symmetry-guaranteed anti-band crossing and band crossing. However, we point out that these are also consistent with the topological conditions. Since the annihilation process does not break chiral symmetry and hence vorticity is conserved all the way. For $k_\mathrm{W}^2<0$, the gapped phase in Model A assures $\nu_{\mathrm{A}}=0$, and the nodal ring in Model B piercing through the plane perpendicular to the mirror plane accounts for $\nu_{\mathrm{B}}=2$.
We remark that the annihilation of WNs into a nodal ring or not by collisions is consistent with the newly discovered conversion rule \cite{PRL.121.106402}. 
 Based on this idea, we can predict that when the mirror symmetry is broken by perturbation and if chiral symmetry is preserved, two WNs in Model A can still annihilate into an insulating phase, while WNs in Model B can collide but not annihilate.

%\section{Summary and discussion}
\emph{Summary.}---
% ------------- %
%  summary		%
% ------------- %
We have clarified that the chiral LLs need the topological protection from vorticity instead of chirality. With a strong magnetic field, two neighbor WNs coalesce into Landau orbits that possess finite- or zero-energy zeroth LLs depending on the net vorticity being 0 or 2 in the plane perpendicular to the field.
The persistence of zero energies with vorticity 2 is robust and has conceptual appeal in searching topologically protected states.
This finding reveals that to characterize topology of a WN chirality as well as vorticity are required in certain cases.
Moreover, we also demonstrate that the net vorticity of a pair of WNs affects impurity scattering, surface states, and the way of WN annihilation.

\begin{acknowledgments}
S.M.H. is supported by the Ministry of Science and Technology (MoST) in Taiwan under grant No. 105-2112-M-110-014-MY3. We thank G. Chang for helpful comments.
\end{acknowledgments}

% ---------- %
%	appendix	%
% ---------- %
%\include{Appendix}

% ------------- %
%   reference	%
% ------------- %
\bibliographystyle{apsrev4-1} % Tell bibtex which bibliography style
%\nocite{*}
%\bibliography{WS_paper}
\bibliography{LL_of_2WS_arxiv}

%\bibliography{LL_of_2WS_arxiv.bbl}

%\documentclass[aps,floatfix,prl, superscriptaddress]{revtex4}
%\documentclass[aps,prl,twocolumn,floatfix, superscriptaddress]{revtex4}
%\documentclass[aps,prl,twocolumn, superscriptaddress]{revtex4-1}
%\documentclass[aps,prb,twocolumn, groupedaddress]{revtex4-1}

%\usepackage{color}
%\usepackage{xcolor}
%\usepackage{amsmath}
%\usepackage{amssymb}
%\usepackage{subfigure}
%\usepackage{graphicx}

%\newcommand\x{\times}
%\setcounter{tocdepth}{1}

%\begin{document}
%\title{Supplementary Material for: \\
%Topologically distinct Weyl-fermion pairs: on the effect of magnetic tunnelling}

%\author{Ming-Chien Hsu}
%\affiliation{Department of Physics, National Sun Yat-sen University, Kaohsiung 80424, Taiwan}	
										  
%\author{Hsin Lin}
%\affiliation{Institute of Physics, Academia Sinica, Nankang Taipei 11529, Taiwan}

%\author{M.\ Zahid Hasan}				
%\affiliation{Laboratory for Topological Quantum Matter and Spectroscopy (B7), 
%Department of Physics, Princeton University, Princeton, New Jersey 08544, USA}
			
%\author{Shin-Ming Huang}
%\email[Corresponding author: ]{shinming@mail.nsysu.edu.tw}
%\affiliation{Department of Physics, National Sun Yat-sen University, Kaohsiung 80424, Taiwan}

%\date{\today}
%\maketitle
%\begin{widetext}
\tableofcontents
%\end{widetext}

\section{Algorithm for finding Landau spectrum}
% ------------ %
%  SVD method  %
% ------------ %
To solve the Hamiltonian with variable $q$ and its derivative $\partial/\partial q$, we can use the language of raising and lowering operators. The replacement is
\[	\sqrt{\frac{2}{g}}q   = \frac{a+a^\dagger}{\sqrt{2}}, \text{ and } 
    \sqrt{\frac{g}{2}}\frac{\partial}{\partial q} = \frac{a-a^\dagger}{\sqrt{2}}
\]
which guarantees the $[a, a^\dagger] = 1$. By imposing $a|0\rangle = 0$, all the necessary relations are then found as $a|n\rangle = \sqrt{n}|n-1\rangle$, $a^\dagger|n\rangle= \sqrt{n+1}|n+1\rangle$ and $a^\dagger a|n\rangle = n\rangle$, where $n=0, \cdots, L, \cdots$ labels the basis with well defined particle numbers.
The eigen-differential problem is then converted into a matrix problem, and we can numerically solve the Landau spectrum of the system by matrix diagonalization. Since numerically we always solve it with a finite matrix of size $L$, the truncation of the operators $a$ and $a^\dagger$ always break canonical commutation relations from the highest few levels. For operators of order $k$, such as $a^\dagger a^\dagger a^\dagger$ of order 3, the levels which break the commutation relations happen at the highest $k$ levels.
Especially, when written in the basis of $|0\rangle, \dots, |L-1\rangle$, the form of $a^\dagger$ is of the $L\times L$ matrix as
\[ a^\dagger = \left( \begin{array}{ccccc}
				0 	   &  	 0    &  0 		& \dots  	 & 0    \\
				1 	   &	 0 	  &  0 		& \dots  	 & 0	\\
				0	   & \sqrt{2} &  0		& \dots  	 & 0	\\
				\vdots & \vdots	  &	\ddots  & \ddots	 & 0	\\
				0	   &	 0	  &	\dots	& \sqrt{L-1} & 0	\\
				\end{array}
			\right)
\]
The highest $L$ basis breaks the $[a, a^\dagger] = 1$ and produces pseudo zero eigenvalues from the state of $[0, 0, \dots, 0, 1]^T$. Therefore, no matter how large the matrix is used, there always would be pseudo zero or near zero states from the highest few particle number basis. This causes serious problems since what we concern is the low level states near zero energies. Some algorithm may use the regularization by adding some big numbers at the highest fewest levels to reduce their contribution. However, this is inconvenient for our case since we do not know how many zero energy states exist {\it a priori}. For two or multiple zero eigen energy solutions, linear combination from these eigenstates is always allowed and we do not have good rules to rule out pseudo solutions without ruining the true solutions. 

Since what we concern is only the low-lying Landau spectrum near the Fermi level 0, we know their contributions all come from the low particle number bases. We then develop an efficient algorithm to exclude the pseudo solutions. For operators of order $k$, the pseudo states come from the highest $k$ basis, and we can restrict the solutions to be in the basis of $|0\rangle, \dots, |L-k\rangle$ of the Hilbert space. This can be effectively achieved by truncating the operator of $L\times L$ matrix into matrix of $L\times (L-k)$. The full Hamiltonian of size of $2L\times 2L$ then becomes matrix of size of $2L\times 2(L-k)$. Using the singular value decomposition (SVD) factorization, we can find the eigen-energies of system without contamination from high lying states. In our case of Hamiltonian which is at most of order 3, we always drop the last 4 bases, namely choosing $k=4$.

For the $m\times n$ matrix $M$, there exists the SVD factorization to be of the form of $M = U\Sigma V^\dagger$, where $U$ is an $m\times m$ unitary matrix whose columns are called the left-singular vectors of $M$, $V$ is an $n\times n$ unitary matrix with columns called right-singular vectors of $M$, and $\Sigma$ is a diagonal $m\times n$ matrix with non-negative real numbers on the diagonal.
The right-singular vectors of $M$ are a set of orthonormal eigenvectors of $M^\dagger M$. 
 For our purpose, the right-singular matrix $V$ serves to find the eigenstates and thus determines the eigen-energies. Below we will demonstrate how to find the low energy spectrum of interest.

Suppose the low energy eigenstates for the Hamiltonian $H$ has the support at most up to $L_k$, namely the mixture components from Landau levels than $L_k$ are zero. In the following SVD approach to get rid of pseudo solutions, we must guarantee that $L-k > L_k$. This can be always be achieved since $L_k$ is usually not very large and we can choose large enough $L$ to guarantee this requirement. The number of truncated columns $k$ can be chosen to be small, say $k=4$ for the system of order 3. Then we can write down the eigenstate of interests to be in the form of
\begin{equation}
  \psi = \left(	\begin{array}{c}
		\tilde{\phi} \\ 0  \\  \vdots  \\ 0  \\
	    \tilde{\chi} \\ 0  \\  \vdots  \\ 0  
	    \end{array}   \right),
\label{Eq:psi_truncate}
\end{equation}
where the $(0, \dots, 0)^T$ are located at the last $k$ Landau levels to be truncated at the up and down spin space separately. 
Both $\tilde{\phi}$ and $\tilde{\chi}$ are columns of size $(L-k)\times 1$ and have support up to $L_k$. Therefore, the weighting components of the $\{L_k+1, \cdots, L-k\}$ levels for them are actually zero. 
Assume that $k=2$ in the following sketch of proof, and then the finite size Hamiltonian $H$ of $2L\times 2L$ matrix can be written as
\[
  H = \left( \begin{array}{ccccc|ccccc}
    & & & \x & \x & & & & \x & \x		\\
    \multicolumn{3}{c}
      {\scalebox{1.7}{$\tilde{H}_{11}$} 
      }
      	 & \vdots & \vdots & 
    \multicolumn{3}{c}
      {\scalebox{1.7}{$\tilde{H}_{12}$} 
      } 	  
		 & \vdots & \vdots \\	   
    & & & \x & \x & & & & \x & \x   \\    \hline
    & & & \x & \x & & & & \x & \x	\\
    \multicolumn{3}{c}
      {\scalebox{1.7}{$\tilde{H}_{21}$} 
      }
      	 & \vdots & \vdots & 
    \multicolumn{3}{c}
      {\scalebox{1.7}{$\tilde{H}_{22}$} 
      } 	  
		 & \vdots & \vdots \\	   
    & & & \x & \x & & & & \x & \x  \\     
  \end{array}	\right),
\]
where the $2k$ columns filled with $\x$ denotes the columns to be truncated. For the eigenstate $\psi$, we can have $H\psi = E\psi$, in which we are interested in low energy $E$ regime. Since components of the last $\{L-k+1, \dots, L\}$ levels for the eigenstates of interest are zero, The columns with $\x$ for the Hamiltonian actually have no effect. We can then drop them and collect the truncated Hamiltonian denoted as $\tilde{H}$ to be  
\[  \tilde{H} = \left(  \begin{array}{c|c}
		\tilde{H}_{11}	&   \tilde{H}_{12}	\\		\hline
		\tilde{H}_{21}	&	\tilde{H}_{22}	\\
	\end{array}		\right),
\]
which is $2L\times 2(L-k)$ matrix. We then do the SVD factorization to have $\tilde{H} = U\Sigma V^\dagger$, with each column vector of $V$ denoted as 
\[	\tilde{\psi} = \left(\begin{array}{c}
				\tilde{\phi}  \\    \tilde{\chi}
				\end{array}		\right)
\]
which is a column of size $2(L-k)\times 1$. Padding with zeros in the form of Eq.\ (\ref{Eq:psi_truncate}) for $\tilde{\psi}$ to become $\psi$, we can have $\psi$ as the eigenstate of $H^2$ with eigenvalue of $E^2$. Doing some linear combination of eigenstates with the same eigenvalue $\lambda = E^2$, we obtain the eigenstate $\psi'$ satisfying  the eigenequation $H\psi' = E\psi'$. If such eigenstate cannot be found, it means that the eigenstate with eigenvalue $E$ for the system has components mixing from Landau levels no smaller than $L$ such that the form of Eq.\ (\ref{Eq:psi_truncate}) with the chosen size $L$ cannot the eigensolutions of $H$. In such case, we increase the numerical system size $L$ until the eigensolutions can be found for the low energy regime. 

The reason that the $\psi$ constructed from $\tilde{\psi}$ from the SVD can be eigenstate of $H^2$ is simple. The Hermitian conjugate of the Hamiltonian $H$ is
\[
  H^\dagger = \left( \begin{array}{cccc|cccc}
    & & & & & & & 		\\
    \multicolumn{4}{c|}
      {\raisebox{.5\normalbaselineskip}{ \scalebox{1.7}{$\tilde{H}_{11}^\dagger$} }
      }				&
    \multicolumn{4}{c}
      {\raisebox{.5\normalbaselineskip}{ \scalebox{1.7}{$\tilde{H}_{12}^\dagger$} }
      } 	  		\\	   
    \x & \cdots & \cdots & \x & \x & \cdots & \cdots & \x   \\   
    \x & \cdots & \cdots & \x & \x & \cdots & \cdots & \x   \\    \hline    
    & & & & & & & 		\\
    \multicolumn{4}{c|}
      {\raisebox{.5\normalbaselineskip}{ \scalebox{1.7}{$\tilde{H}_{21}^\dagger$} }
      }				&
    \multicolumn{4}{c}
      {\raisebox{.5\normalbaselineskip}{ \scalebox{1.7}{$\tilde{H}_{22}^\dagger$} }
      } 	  		\\   
    \x & \cdots & \cdots & \x & \x & \cdots & \cdots & \x  \\     
    \x & \cdots & \cdots & \x & \x & \cdots & \cdots & \x  \\         
  \end{array}	\right),
\]
such that 
\[
  H^\dagger H = \left( \begin{array}{ccccc|ccccc}
    & & & \x & \x & & & & \x & \x 		\\
    \multicolumn{3}{c}
      {\raisebox{.5\normalbaselineskip}{ \scalebox{1.7}{$B_{11}$} }
      }				
      	&  \vdots  &  \vdots  &
    \multicolumn{3}{c}
      {\raisebox{.5\normalbaselineskip}{ \scalebox{1.7}{$B_{12}$} }
      } 	
      	&  \vdots  &  \vdots		\\	   
    \x & \cdots & \x & \x & \x & \x & \cdots & \x & \x  & \x  \\   
    \x & \cdots & \x & \x & \x & \x & \cdots & \x & \x  & \x  \\    \hline    
    & & & \x & \x & & & & \x & \x 		\\    
    \multicolumn{3}{c}
      {\raisebox{.5\normalbaselineskip}{ \scalebox{1.7}{$B_{21}$} }
      }				
      	&  \vdots  &  \vdots  &
    \multicolumn{3}{c}
      {\raisebox{.5\normalbaselineskip}{ \scalebox{1.7}{$B_{22}$} }
      } 	  	
      	&  \vdots  &  \vdots        \\
    \x & \cdots & \x & \x & \x & \x & \cdots & \x & \x  & \x  \\         	
    \x & \cdots & \x & \x & \x & \x & \cdots & \x & \x  & \x  \\         	
  \end{array}	\right),
\]
where 
\[	B \equiv   \tilde{H}^\dagger \tilde{H} = \left(	\begin{array}{cc}
			B_{11}	&	B_{12}	\\
			B_{21}	&	B_{22}	\\
			\end{array}		\right).
\]
The $\tilde{\psi}$ obtained from SVD of $\tilde{H}$ is just the eigenstates of $\tilde{H}^\dagger \tilde{H}$. As long as the eigenstate $\psi$ satisfying $H\psi = E\psi$ has the support $L_k < L-k$, namely components from highest $k$ levels are zero, the form of Eq.\ (\ref{Eq:psi_truncate}) constructed from $\tilde{\psi}$ would fall into eigen-solutions of $H^2\psi = H^\dagger H\psi = E^2\psi$. 

\section{The analytical solutions for Model B with $\tilde{\alpha}=0$}
The spectrum of Model B with $\tilde{\alpha}=0$ for small $g$ can also be found. These solutions are also plotted as red dots in Fig.~1 (b) in the main text for comparison. Model B under field defined in the main text and is rewritten here
\begin{equation}
\begin{split}
 H_{\mathrm{B}}^{\prime} = \frac{\omega_{c}^2}{4E_{\mathrm{VH}}} 
	 \left\{ 
	 \left[\frac{1}{g}(q^2-1) 
	  + g\frac{\tilde{\alpha}}{4}\frac{\partial^2}{\partial q^2}
	  -  g\frac{\tilde{\alpha}}{4} q_z^2         \right] \sigma_x	
	  \right. \\ \left. 
       + i \left(q\frac{\partial}{\partial q} + \frac{1}{2} \right)\sigma_y 
   	  + q_z q\sigma_z  \right\},
\label{Eq:B1app}
\end{split}
\end{equation}
where the prime stands for a system under a magnetic field. 
 
  In the following we present the derivations for the solutions when $\tilde{\alpha}=0$ and $k_z=0$ for Model B. The Hamiltonian to solve is $H^\prime_{\mathrm{B}}$ in Eq. (\ref{Eq:B1app}) with $\tilde{\alpha}=0$ and $q_z=0$, and we assume the eigenstate to be $(\chi(q), \Phi(q))^T$.
By squaring $H_\mathrm{B}^\prime$, we can decouple $\chi$ and $\Phi$ as
\begin{equation}
	\begin{split}
\left\{ -\left(q\frac{\partial}{\partial q} + \frac{1}{2}\right) 
							\left(q\frac{\partial}{\partial q} + \frac{1}{2}\right) 
						+ \frac{2}{g} q^2
				 	    + \frac{1}{g^2}(q^2 - 1)^2  \right\} \chi(q)         \\
	= \varepsilon^2 \chi(q)	\\
\left\{ -\left(q\frac{\partial}{\partial q} + \frac{1}{2}\right)
							\left(q\frac{\partial}{\partial q} + \frac{1}{2}\right)
						- \frac{2}{g} q^2
						+ \frac{1}{g^2}(q^2 - 1)^2 \right\} \Phi(q) 		\\
	= \varepsilon^2 \Phi(q) 
	\end{split}
\end{equation}
where the energy is rescaled to a dimensionless quantity defined as $\varepsilon= E/ \frac{\omega_c^2}{4E_\mathrm{VH}}$. Owing to
\begin{equation}
\begin{split}
\left(q\frac{\partial}{\partial q} + \frac{1}{2}\right) \frac{\chi}{\sqrt{q}} 
	= \frac{1}{\sqrt{q}} \left( q \frac{\partial}{\partial q} \right)\chi, & \text{ and thus }  \\
\left(q\frac{\partial}{\partial q} + \frac{1}{2}\right)^2 \frac{\chi}{\sqrt{q}} 
	= \frac{1}{\sqrt{q}} \left( q \frac{\partial}{\partial q} \right)^2 \chi,
\end{split}
\end{equation}
we firstly take $\chi(q) = \frac{1}{\sqrt{q}}e^{-\frac{q^2}{2g}} \tilde{\chi}(q)$ and 
$\Phi(q) = \frac{1}{\sqrt{q}}e^{-\frac{q^2}{2g}} \tilde{\Phi}(q)$ into the differential equations and obtain
\begin{equation}
\begin{array}{l}
\left[ q^2\frac{\partial^2}{\partial q^2} 
		+ \left(1 - \frac{2}{g}q^2 \right) q \frac{\partial}{\partial q}
		+ \frac{2}{g^2} (1 - 2g) q^2 \right] \tilde{\chi}(q) = \lambda^2 \tilde{\chi}(q),	\\
\left[ q^2\frac{\partial^2}{\partial q^2} 
		+ \left(1 - \frac{2}{g}q^2 \right) q \frac{\partial}{\partial q}
		+ \frac{2}{g^2} q^2 \right] \tilde{\Phi}(q) = \lambda^2 \tilde{\Phi}(q),	\\
\end{array}
\end{equation}
where $\lambda^2 = \frac{1}{g^2} - \varepsilon^2$. Note that the exponential factor $e^{-\frac{q^2}{2g}}$ is to eliminate the quartic term $q^4$. Assuming $\lambda > 0$, we continue to take $\tilde{\chi}(q) = q^\lambda f_1(q)$ and $\tilde{\Phi}(q) = q^\lambda f_2(q)$ and have
\begin{equation}
\begin{array}{l}
q^2\frac{\partial^2 f_1}{\partial q^2} 
		+ \left(1 + 2\lambda - \frac{2}{g}q^2 \right) q \frac{\partial f_1}{\partial q}
		+ \frac{2}{g^2} (1 - 2g - g\lambda) q^2 f_1  = 0,	\\
q^2\frac{\partial^2 f_2}{\partial q^2} 
		+ \left(1 + 2\lambda - \frac{2}{g}q^2 \right) q \frac{\partial f_2}{\partial q}
		+ \frac{2}{g^2} (1 - g\lambda) q^2 f_2  = 0.			\\
\end{array}
\end{equation}
Then we rescale the length by $\bar{q} = \frac{q}{\sqrt{g}}$, and obtain
\begin{equation}
\begin{array}{l}
\bar{q}^2\frac{\partial^2 f_1}{\partial \bar{q}^2} 
		+ \left(1 + 2\lambda - 2\bar{q}^2 \right) \bar{q} \frac{\partial f_1}{\partial \bar{q}}
		+ \frac{2}{g} (1 - 2g - g\lambda) \bar{q}^2 f_1  = 0,	\\
\bar{q}^2\frac{\partial^2 f_2}{\partial \bar{q}^2} 
		+ \left(1 + 2\lambda - 2 \bar{q}^2 \right) \bar{q} \frac{\partial f_2}{\partial \bar{q}}
		+ \frac{2}{g} (1 - g\lambda) \bar{q}^2 f_2  = 0	.		\\
\end{array}
\end{equation}
The final step is to change the variable from $\bar{q}$ to $\rho=\bar{q}^2$, which results in
\begin{equation}
\begin{array}{l}
4\rho^2\frac{\partial^2 f_1}{\partial \rho^2} 
		+ 4 (1 + \lambda - \rho ) \rho \frac{\partial f_1}{\partial \rho}
		+ \frac{2}{g} (1 - 2g - g\lambda) \rho f_1  = 0	\\
4\rho^2\frac{\partial^2 f_2}{\partial \rho^2} 
		+ 4 (1 + \lambda - \rho ) \rho \frac{\partial f_2}{\partial \rho}
		+ \frac{2}{g} (1 - g\lambda) \rho f_2  = 0	\\
\end{array}
\end{equation}
In the above, we already used
\begin{equation}
\begin{array}{ccl}
	\bar{q}  \frac{d f}{d\bar{q}}     &=& 2 \rho   \frac{d f}{d \rho},		\\
	\bar{q}^2\frac{d^2 f}{d\bar{q}^2} &=& 4 \rho^2 \frac{d^2 f}{d \rho^2} + 2\rho \frac{d f}{d \rho} .
\end{array} 
\end{equation}
Reformulating the last two equations, we have the so called associated (generalized) Laguerre equations
\begin{equation}
\begin{array}{l}
\rho \frac{\partial^2 f_1}{\partial \rho^2} 
		+ (\lambda +1 - \rho ) \frac{\partial f_1}{\partial \rho}
		+ (n -1) f_1  = 0,	\\
\rho \frac{\partial^2 f_2}{\partial \rho^2} 
		+ (\lambda +1 - \rho ) \frac{\partial f_2}{\partial \rho}
		+  n f_2  = 0,	\\
\end{array}
\end{equation}
where $n= \frac{1}{2g} - \frac{\lambda}{2} = 0, 1, 2, \cdots$. The solutions $f_1$ and $f_2$ will be the associated Laguerre polynomials: $f_1(\rho) = L^\lambda_{n-1}(\rho)$ and $f_2(\rho) = L^\lambda_n(\rho)$ with $n=0, 1, 2, \cdots$. For $n=0$, we have to take $f_1=0$. As 
$n= \frac{1}{2g} - \frac{1}{2}\sqrt{\frac{1}{g^2} - \varepsilon^2} \geq 0$ is a non-negative integer, the energy eigenvalues are 
\begin{equation}
E_n = 2 \frac{\omega_c^2}{4E_\mathrm{VH}}\left[ n \left(\frac{1}{g} - n \right) \right]^{\frac{1}{2}}.
\end{equation}

Here $E_n\leq \frac{1}{g}$ as $\lambda^2 = \frac{1}{g^2} - \varepsilon^2 > 0$. 
The relevant state is the lowest Landau level $n=0$ which gives the zero-energy state $E_{n=0}=0$. 
It's eigenstate are thus $(0, \Phi)^T$ with 
\begin{equation}
\Phi(q) = \frac{1}{\sqrt{q}} e^{-\frac{q^2}{2g}} q^\lambda L^\lambda_n \left(\frac{q^2}{g} \right)
	    = q^{\lambda-\frac{1}{2}} L^\lambda_n \left(\frac{q^2}{g} \right)  e^{-\frac{q^2}{2g}},
\end{equation} 
where $L^\lambda_n$ is a polynomial function of $q^2$ to degree $n$. Therefore, the normalizability demands $\lambda \geq \frac{1}{2}$, that is, 
$0 \leq n \leq \frac{1}{2}\left(\frac{1}{g} - \frac{1}{2} \right)$. 
The missing states for large $n$ not satisfying the constraint become extended states whose spectra are continuous. This is an artefact of this model with $\alpha=0$ which omit the $k_y$ and $k_z$ dependence in the $\sigma_x$ term, since it produces an open equienergy contour for $E\geq E_{\mathrm{VH}}$ and hence the cyclotron orbit is not confined.

\section{The analytical form of zero energy solutions for Model $H_{\mathrm{B}}$}
% ----------------------------- %
%	WF of the zero Landau level	%
% ----------------------------- %
Here we demonstrate the analytical form of the zero energy solutions for Model B with $k_z=0$. The zero-eigenvalue problem is $ H_{\mathrm{B}}^{\prime}  \Psi =0$, where $H_{\mathrm{B}}^{\prime}  $ is an $2\times2$ off-diagonal matrix with elements 
\begin{equation}
\frac{1}{g}(q^2-1) + g\frac{\tilde{\alpha}}{4}\frac{\partial^2}{\partial q^2} \pm \left(q\frac{\partial}{\partial q} + \frac{1}{2} \right).
\end{equation}
The prime means the system in a magnetic field. We could try solutions either as $\Psi =\left(0, \psi\right)^{T}$ or $\Psi =\left(\psi, 0 \right)^{T}$, but it turns out that the second choice is not normalizable. Then  we are going to solve the differential equation
\begin{equation}
\left[ g\frac{\tilde{\alpha}}{4}\frac{\partial^2}{\partial q^2} 
    +   \left(q\frac{\partial}{\partial q} + \frac{1}{2} \right) + \frac{1}{g}(q^2-1) \right]\psi(q)=0.
    \label{EVproblem}
\end{equation}
Its large-$q$ limit can be conquered by setting $\psi(q) = e^{-\kappa q^2}\Phi(q)$ and taking it into Eq. (\ref{EVproblem}) to obtain $\kappa$. As $\kappa>0$ for normalizability, we have $\kappa=\frac{1+\sqrt{1-\tilde{\alpha}}}{g\tilde{\alpha}}$. The other choice is $\kappa=\frac{1-\sqrt{1-\tilde{\alpha}}}{g\tilde{\alpha}}$, but this would lead to additional exponential term from modified Hermite differential equation later. After absorbing the additional exponential term, this results in the same result from $\kappa=\frac{1+\sqrt{1-\tilde{\alpha}}}{g\tilde{\alpha}}$ and therefore we focus on the plus sign choice.
It follows the differential equation for $\Phi(q)$, which is
\begin{equation}
+\frac{g\tilde{\alpha}}{4}\frac{\partial^2\Phi}{\partial q^2} -\sqrt{1-\tilde{\alpha}}q\frac{\partial\Phi}{\partial q} -[\frac{1}{g}+\frac{\sqrt{1-\tilde{\alpha}}}{2}]\Phi = 0.
\end{equation}
By defining
$\xi=\sqrt{\frac{2\sqrt{1-\tilde{\alpha}}}{g\tilde{\alpha}}}$, $\lambda = \frac{2+g\sqrt{1-\tilde{\alpha}}}{-g\sqrt{1-\tilde{\alpha}}}$ and $\theta=\xi q$, the equation is then transformed into the Hermite differential equation
\begin{equation}
\frac{\partial^2 \Phi}{\partial\theta^2} -2\theta\frac{\partial\Phi}{\partial\theta} + \lambda \Phi =0.
\end{equation}
The solution for $\Phi$ is 
${_1F_1}(-\frac{1}{4}\lambda; \frac{1}{2}; \theta^2)$ and $H_{\lambda/2}(\theta))$, where $_1F_1(a;b;x)$ and $H_\mu(x)$ are the Kummer confluent hypergeometric function and the Hermite polynomial respectively.

It is known that $H_{\lambda/2}(\xi q)$ is not purely even or odd with respect to $q$. As one can find that the even part of $H_{\lambda/2}(\xi q)$ is actually $_1F_1(-\frac{1}{4}\lambda; \frac{1}{2}; \xi^2 q^2)$, it is consistent that after reconstruction the zero-energy eigenfunctions are either of even or odd parity in $q$. 

\begin{figure}[ht]
%\begin{minipage}[b]{.45\textwidth}
\begin{center}
	\includegraphics[width=0.5\textwidth]{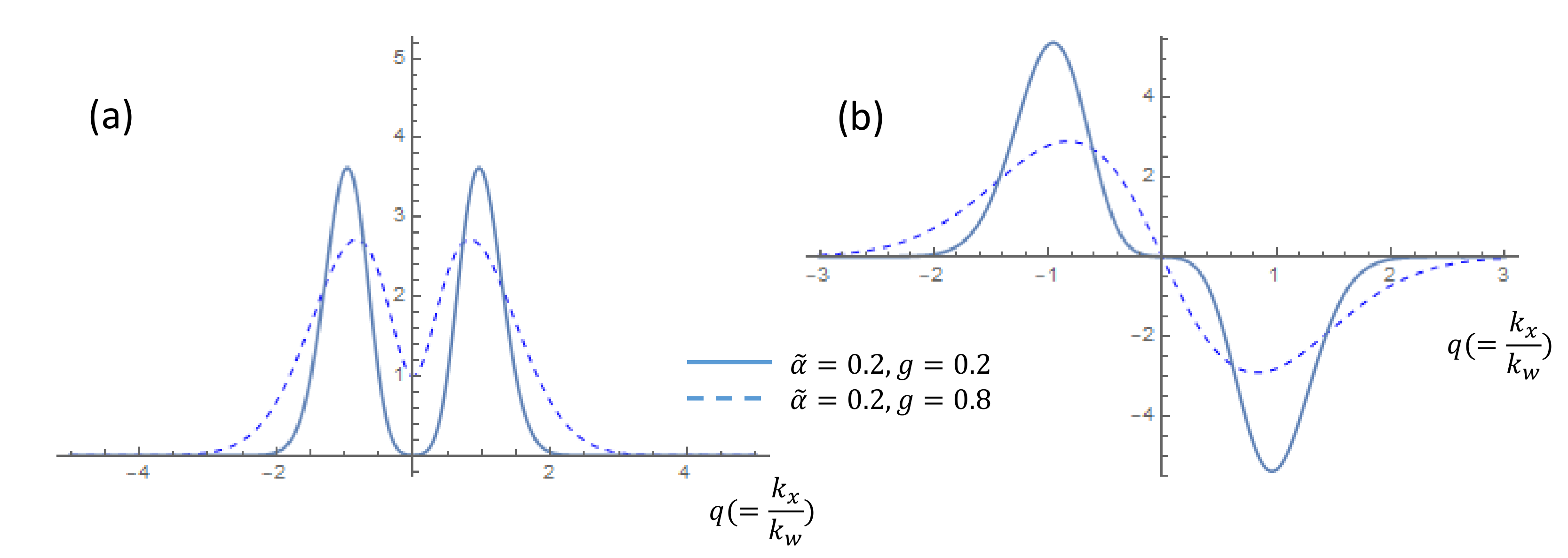}
\caption{
The zero energy eigen-functions in the (a) q-even solutions and (b) q-odd solutions. The parameters of $g=0.2$ and $g=0.8$ both for $\tilde{\alpha}=0.2$ are chosen for comparison.
}
\label{fig:zeroWF}
\end{center}
%\end{minipage}
\end{figure}

The zero-energy eigen-functions are plotted in Fig.\ \ref{fig:zeroWF}. The $g$ value proportional to the field strength controls the coupling between the two nodes. In small $g$, the Landau orbits are well separated and have each center around the Weyl nodes. The orbits start to overlap for larger $g$ so that they tend to move toward the mirror plane.  The trend of tuning $g$ for both $q$-even and $q$-odd solutions are the same.

\section{Models including pairs of Weyl nodes separated in $k_z$}
To study chiral anomaly of pairs of close Weyl nodes (here separated in $k_x$ direction) under the effect of magnetic field along $z$ direction, we must include the other pair of nodes separated in $z$ direction.
By shifting the Weyl nodes in $H_\mathrm{A}$ and $H_\mathrm{B}$ to 
$k_z =  k_{\mathrm{V}}$ and include the other pair of Weyl nodes at 
$k_z = -k_{\mathrm{V}}$, we can have the modified models, denoted as $H_{\tilde{\mathrm{A}}}$ and $H_{\tilde{\mathrm{B}}}$, in the form of
\begin{equation}
\begin{array}{cl}
 H_{\tilde{\mathrm{A}}} =& \frac{1}{2m}(k_x^2 - k_{\mathrm{W}}^2 
               -  \alpha k_\parallel^2)\sigma_x   + v_{\parallel} k_y \sigma_y    \\
 		 &  
		    + \frac{v_\parallel}{2k_{\mathrm{V}}}(k_z^2-k_{\mathrm{V}}^2) \sigma_z,    \\
H_{\tilde{\mathrm{B}}} = &   \frac{1}{2m}(k_x^2   - k_{\mathrm{W}}^2
              - \alpha k_\parallel^2)\sigma_x 	+ \frac{v_\parallel}{k_W} k_x k_y \sigma_y   \\
	 	 &	
 			+ \frac{v_\parallel}{2k_{\mathrm{V}}} \frac{k_x}{k_W} (k_z^2 - k_{\mathrm{V}}^2) \sigma_z.
\end{array}		      
\label{model_H_AB_2Z}
\end{equation}
The Weyl nodes of the first pair are then located at 
$(\pm k_\mathrm{W}, 0,  k_{\mathrm{V}})$, while the other pair is located at $(\pm k_\mathrm{W}, 0, -k_{\mathrm{V}})$.
 Since we are looking at physics near the Weyl nodes and their low energy spectrum, the $\alpha k_\parallel^2$ term does not play much role. In most of the time they can be even dropped. The value of $\alpha$ under discussion is therefore small, and the $\alpha k_\parallel^2$ term affects mainly the dispersion in higher energy and does not influence the low energy of interest much. 
 
Here Model $\tilde{A}$ and Model $\tilde{B}$ still have the mirror plane $k_x=0$,
 and we do not put in additional symmetry relation between the first pair of WNs and the second pair for simplicity. In this way, the effect of different choices of $M_x= \sigma_0$ or $\sigma_x$ can be clearly seen. 
Besides, usually in real materials, additional WNs are far away from the pair of interest such that their effect can be discarded since they are far from reach of the magnetic length scale under reasonable field strength.   
Therefore, for simplicity we only compare one mirror plane with different operator choices in order to elucidate the symmetry impacts. 

%The Fermi velocities at the first node pair $(\pm k_\mathrm{W}, 0, k_{\mathrm{V}z})$ are 
%$(\pm\frac{k_{\mathrm{W}}}{m}, v_y, v_z)$ for $\tilde{H}_\mathrm{A}$ and 
%$\pm (\frac{k_{\mathrm{W}}}{m}, v_y, v_z)$ for $\tilde{H}_\mathrm{B}$. 
%Similarly, the Fermi velocities at $(\pm k_\mathrm{W}, 0, -k_{\mathrm{V}z})$ are 
%$(\pm\frac{k_{\mathrm{W}}}{m}, v_y, -v_z)$ for $\tilde{H}_\mathrm{A}$ and 
%$\pm (\frac{k_{\mathrm{W}}}{m}, v_y, -v_z)$ for $\tilde{H}_\mathrm{B}$. 

In the usual Weyl semimetal, the separation of WNs in the $k_z$ direction is larger than the $k_{\mathrm{W}}$, i.e.  $k_{\mathrm{V}} > k_{\mathrm{W}}$. 
Due to the other pair of nodes, the term of $q_z \sigma_z$ in Model A Hamiltonian under magnetic field $H_{\mathrm{A}}^{\prime}$ is replaced by 
 $\frac{1}{g}A_z(q_z^2 - 1)$ 
in $H_{\tilde{\mathrm{A}}}^\prime$ under field while the term $q_z q \sigma_z$ in $H_{\mathrm{B}}^{\prime}$ is replaced by $\frac{1}{g}A_z q(q_z^2 - 1)$
 in $H_{\tilde{\mathrm{B}}}^\prime$, where 
$A_z = g\frac{v_\parallel k_{\mathrm{V}}}{2} \left( \frac{\omega_c^2}{4E_\mathrm{VH}}\right)^{-1} = (\frac{v_\parallel}{v_x}) (\frac{k_\mathrm{V}}{k_\mathrm{W}})$
 with the new definition of the dimensionless $q_z=k_z/k_{\mathrm{V}}$. 
The $-\frac{\alpha}{2m} k_\parallel^2$ term in both models $H_{\mathrm{A}}^\prime$ and $H_{\mathrm{B}}^\prime$ would correspondingly become
$\frac{\omega_c^2}{4 E_\mathrm{VH}} \left[  
g \frac{\tilde{\alpha}_y}{4}\frac{\partial^2}{\partial q^2}
 			  - \frac{1}{g}\tilde{\alpha}_z q^2_z  \right]$,
where $\tilde{\alpha}_y = \alpha(\frac{v_x}{v_\parallel})^2$,
  $\tilde{\alpha}_z = \alpha (\frac{k_\mathrm{V}}{k_\mathrm{W}})^2$.
Since $q$ and $q_z$ are independent to each other, $q_z$ as a good quantum number can be treated as a parameter and the Hamiltonian is solved at fixed $q_z$ each time. 

Explicitly, the Hamiltonian under field to solve for Model $\tilde{A}$ is then
\begin{equation}
\begin{split}
 H_{\tilde{\mathrm{A}}}^{\prime} = \frac{\omega_c^2}{4 E_\mathrm{VH}}  \left\{
	    \left[\frac{1}{g}(q^2-1) + g \frac{\tilde{\alpha}_y}{4}\frac{\partial^2}{\partial q^2}
 			  - \frac{1}{g}\tilde{\alpha}_z q^2_z  	 	 \right]\sigma_x	 \right. \\ \left. 
	 +  i \frac{\partial}{\partial q}  \sigma_y	
	 +        \frac{1}{g} A_z (q^2_z -1) 	 \sigma_z
   		 \right\},
\end{split}
\end{equation}
while the Hamiltonian for Model $\tilde{B}$ is then
\begin{equation}
\begin{split}
 H_{\tilde{\mathrm{B}}}^{\prime} = \frac{\omega_c^2}{4 E_\mathrm{VH}}  \left\{
	    \left[\frac{1}{g}(q^2-1) 
	     + g \frac{\tilde{\alpha}_y}{4}\frac{\partial^2}{\partial q^2}
 			  - \frac{1}{g}\tilde{\alpha}_z q^2_z  	 	 \right]\sigma_x	\right. \\ \left. 
	 +   i \left(q\frac{\partial}{\partial q}  + \frac{1}{2} \right)  \sigma_y	
	 +        \frac{1}{g} A_z q(q^2_z -1) 	 \sigma_z
   		 \right\}
\end{split}
\end{equation}

\section{Solutions for rotated magnetic field in the yz plane}
\begin{figure}[ht]
%\begin{minipage}[b]{.45\textwidth}
\begin{center}
	\includegraphics[width=0.45\textwidth]{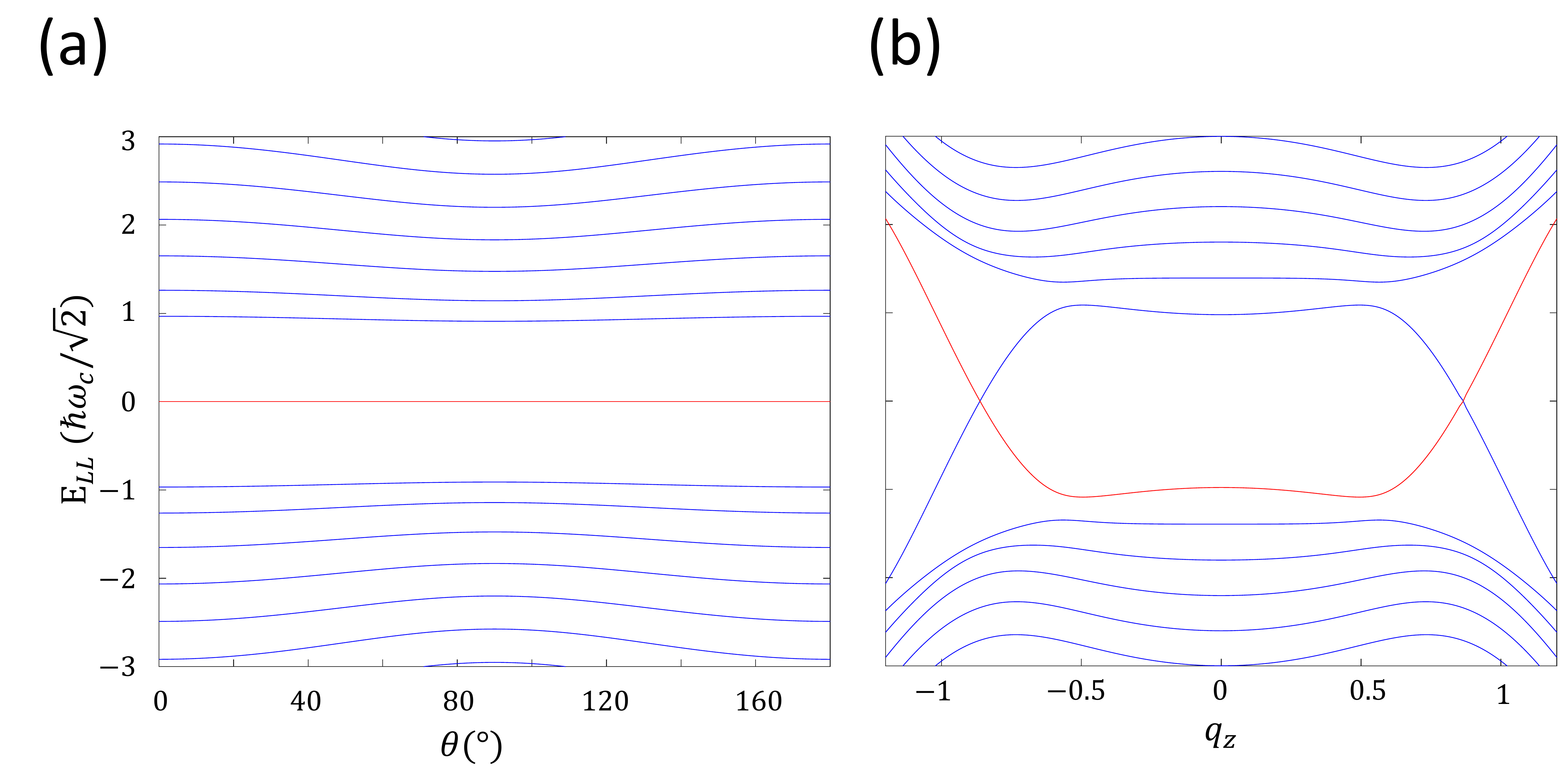}
\caption{(a) The case of $\alpha=0.05$, $v_y/ v_x = 0.5$, and $v_z/v_x=0.4$ so that  $\tilde{\alpha}_y = 0.2$ and $\tilde{\alpha}_z= 0.3125$ for the model Hamiltonian $H^{\prime}_\mathrm{B}$ is presented. With fixed value of $g=0.8$ and $k'_z=0$, the zero Landau energies persist in all rotated angles $\theta$.
(b) The $k_z/k_{\mathrm{V}}$ dispersion along the field direction is presented for the Hamiltonian $H_{\tilde{\mathrm{B}}}^{\prime}$ with parameters to be
$\alpha=0.05$, $v_y/v_x = v_z/v_x = 0.5$, $k_\mathrm{W}/k_\mathrm{V}=0.2$ . The field strength and direction are chosen to have $g = 0.8$ and rotated angle $\theta=30^\circ$.
}
\label{fig:LLrotateB}
\end{center}
%\end{minipage}
\end{figure}

% ----------------------- %
%   rotated theta result     %
% ----------------------- %
Here we consider the magnetic field rotated in the yz plane which is still perpendicular to the two nodes separated in the $k_x$ direction. Since we mainly concern about whether the chiral Landau levels and the zero energy levels can maintain, we focus on the case following discussions of model $H_\mathrm{B}$ and $H_{\tilde{\mathrm{B}}}$.
The coefficients for $k_y$ and $k_z$ in the Hamiltonian can be different but the physics is the same. For general purpose, we can always rescale $k_y$ and $k_z$ such that $\alpha k_\parallel^2$ have the same coefficients for $k_y$ and $k_z$ while linear terms $\sim v_y k_x k_y \sigma_y$ and $\sim v_z k_x k_z \sigma_z$ have different parallel velocities $v_y$ and $v_z$ for $k_y$ and $k_z$ respectively. Note that the $k_x=0$ plane still need to be dispersional for all $(k_y, k_z)$ to ensure close energy contours such that the magnetic orbits can be formed under all rotated field directions. In the Hamiltonian, this means that $\alpha$ can be small but cannot be zero.
For single pair of Weyl nodes, the Hamiltonian is 
\begin{equation}
 H_\mathrm{B} = \frac{1}{2m}[k_x^2 - k_{\mathrm{W}}^2 
           - \alpha k_\parallel^2]\sigma_x   
     + \frac{v_y}{k_W} k_x k_y \sigma_y 
       + \frac{v_z}{k_W} k_x k_z \sigma_z,
\label{model_H_B3}
\end{equation}
which is already defined in the main text.
The coefficient $\alpha$ mainly influence dispersion in higher energies. Usually $\alpha$ is small, and therefore does not change the low energy spectrum much, including the zero energy levels of concern.
The simpler way to deal with rotated field is to define new momentum coordinates $k'_y$ and $k'_z$ where the new $\hat{z}$ direction is along the field. Suppose the field $\mathbf{B}=B(\sin\theta\hat{y}+\cos\theta\hat{z})$, then the new momentum coordinates are defined as $k'_y=\cos\theta k_y - \sin\theta k_z$ and $k'_z=\sin\theta k_y + \cos\theta k_z$, and still $k_\parallel^2 = k_y^2 + k_z^2 = {k'_y}^2 + {k'_z}^2$.
 In such definition, we take the advantage of $k'_z$ still being a good quantum number and $k_x k'_y \rightarrow l_B^{-2}(k_x\bar{x}+\frac{i}{2})$ similar to model $H_\mathrm{B}$ with modified guiding center $x_0=l^2_B k'_y$. 
With definition of 
$\sigma'_y = \cos\theta\sigma_y - \sin\theta\sigma_z$ and 
$\sigma_z' = \sin\theta\sigma_z + \cos\theta\sigma_z$, the Hamiltonian in new coordinates can be written as
\begin{equation}
\begin{array}{cl}
 H_\mathrm{B} =& \frac{1}{2m}[k_x^2 - k_{\mathrm{W}}^2 
                    - \alpha {k_\parallel}^2 ]\sigma_x  + \frac{v_y}{k_\mathrm{W}} k_x k'_y \sigma_y'   \\
    & 
       + \frac{v_z}{k_\mathrm{W}} k_x k'_z \sigma'_z.
\end{array}
\label{model_H_B3rotate}
\end{equation}
The dimensionless momentum in the field direction is defined as 
$q'_z = v_z k'_z \left( \frac{\omega_c^2}{4 E_\mathrm{VH}} \right)^{-1}
= (\frac{2}{g})(\frac{v_z}{v_x})(\frac{k'_z}{k_\mathrm{W}})$.
The dimensionless quantities now are $\tilde{\alpha}_y = \alpha (\frac{v_x}{v_y})^2$ and $\tilde{\alpha}_z = \alpha (\frac{v_x}{v_z})^2$.
Therefore, the Hamiltonian under field in rotated coordinate is then 
\begin{equation}
\begin{split}
 H^{\prime}_\mathrm{B} = \frac{ \omega_c^2}{4 E_\mathrm{VH}} \left\{
	 \left[\frac{1}{g}(q^2-1) 
	 + g\frac{\tilde{\alpha}_y}{4}\frac{\partial^2}{\partial q^2}
 	 - g \frac{\tilde{\alpha}_z}{4}{q^\prime_z}^2  \right]\sigma_x		\right. \\ \left. 
      + i \left(q\frac{\partial}{\partial q} + \frac{1}{2} \right) \sigma'_y
   	  + q'_z q \sigma_z^\prime	 \right\},
\label{Eq:B3Landau}
\end{split}
\end{equation}
where the prime on the left side refers to the Hamiltonian under magnetic field. The independent parameters are  $\tilde{\alpha}$, $g$, and rotated angle $\theta$. Among them, $\tilde{\alpha}$ and $g$ are related to materials properties, i.e. Weyl nodes quantities, and $g$ and $\theta$ are related to field amplitude and direction respectively. As an example, we present the case of $\tilde{\alpha}= 0.05$ and with fixed value of $g=0.8$,
 we rotate the angle  $\theta$ of field with respect to $\hat{z}$ axis from $0^\circ$ to $180^\circ$ and present the result in Fig.~\ref{fig:LLrotateB} (a). It can be found that the zero energies persist in all angles $\theta$ in the plane of $k'_z=0$ perpendicular to the rotated field.

To see if the chiral anomaly can remain for rotated field when two pairs of Weyl nodes are located at $k_z=\pm k_{\mathrm{V}}$, the Hamiltonian for Model A and Model B are Eq.\ (\ref{model_H_AB_2Z}), and we focus on discussing Model B.

Similarly, when written in the new coordinate $(k_x, k'_y, k'_z)$ of the rotated frame, Model B Hamiltonian is then
\begin{equation}
\begin{array}{cl}
 H_{\tilde{\mathrm{B}}} =& \frac{1}{2m}[k_x^2 - k_{\mathrm{W}}^2 
                    - \alpha {k_\parallel}^2   ]\sigma_x    						
  + \frac{v_y}{k_\mathrm{W}} k_x 
	             (\cos\theta k'_y + \sin\theta k'_z)\sigma_y	    \\
	& + \frac{v_z}{k_\mathrm{W}} k_x
	             \left( \sin^2\theta{k'_y}^2 + \cos^2\theta{k'_z}^2  \right.     \\
	&		\left.	-2\cos\theta\sin\theta k'_y k'_z - k^2_{\mathrm{V}} \right)\sigma_z.	
\end{array}
\label{model_H_B4rotate}
\end{equation}
Different to $H_\mathrm{B}$, the dimensionless momentum along the field direction is defined as $q_z = k'_z/k_{\mathrm{V}}$. Therefore, the dimensionless parameters involving $k'_z$ would change the dependence from $v_x/v_\parallel$
 to $k_{\mathrm{V}}/k_{\mathrm{W}}$. Explicitly, the dimensionless parameters are defined as follows. 
  $\tilde{\alpha}_y = \alpha(\frac{v_x}{v_y})^2$,
  $\tilde{\alpha}_z = \alpha (\frac{k_\mathrm{V}}{k_\mathrm{W}})^2$, 
$A_z = \left( \frac{v_z}{v_x} \right) (\frac{k_\mathrm{V}}{k_\mathrm{W}})$, 
$A_y = (\frac{v_x v_z}{4 v_y^2} ) (\frac{k_\mathrm{W}}{k_\mathrm{V}})$ and $A_{yz} = \frac{v_z}{2 v_y}$, where the definition of $A_z$ is the same as that in single pair of Wely nodes.
Not all the defined dimensionless parameters are independent. 
 Among them, the independent parameters are chosen to be
$\alpha$, $k_{\mathrm{W}} / k_{\mathrm{V}}$, and $\frac{v_x}{v_y}$. The Hamiltonian under magnetic field to solve is then

\begin{equation}
\begin{split}
 H_{\tilde{\mathrm{B}}}^{\prime} = \frac{\omega_c^2}{4 E_\mathrm{VH}}  \left\{
	    \left[\frac{1}{g}(q^2-1) + g \frac{\tilde{\alpha}_y}{4}\frac{\partial^2}{\partial q^2}
 			  - \frac{1}{g}\tilde{\alpha}_z q^2_z  	 	 \right]\sigma_x	\right. \\ 
	 +  \left[i \left(q\frac{\partial}{\partial q}  + \frac{1}{2} \right)\cos\theta
      		  +\frac{2}{g } A_z (q q_z)\sin\theta  \right]\sigma_y						\\
	 +  \left[-g A_y \sin^2\theta      		  
      		  \left(q\frac{\partial}{\partial q} + 1 \right)
      		  \frac{\partial}{\partial q}						   
      		 +\frac{1}{g} A_z q(q^2_z\cos^2\theta -1)              \right.  \\
        \left.  \left.
      		 -2 i A_{yz} \cos\theta\sin\theta
	      		  \left(q\frac{\partial}{\partial q} + \frac{1}{2} \right) q_z
	   \right] \sigma_z
   	 \right\}.
\label{Eq:B4Landau}
\end{split}
\end{equation}
With the reasonable choice of $\alpha=0.05$, $v_y/v_x = v_z/v_x = 0.5$, and $k_{\mathrm{W}} / k_{\mathrm{V}}=0.2$,
 we present the case of $g = 0.8$ and $\theta=30^\circ$ in Fig.~\ref{fig:LLrotateB} (b).
The corresponding values are $\tilde{\alpha}_y = 0.2$, $\tilde{\alpha}_z= 1.25$, $A_y = 0.1$, $A_z = 2.5$, and $A_{yz} = 0.5$.  
It is found that the chiral anomaly still remains since the chiral Landau levels near each pair of nodes are robust. 
 
% ------------------- %
%   surface states       %
% ------------------- %
\section{surface states}
We are going to solve Weyl semimetal slabs for Model A and Model B. In the $x$ and $y$ directions, the sizes are infinity, while it is semi-infinity in the $z$ direction. Assume that the Weyl semimetal systems are built for $z<0$ adjacent to vacuum for $z>0$. We will analyze Model B first and then Model A since the former model is new to us. 
\subsection{Model B}
To model a vacuum-semimetal interface, we introduce a mass term in the Hamiltonian as
\begin{equation}
H_{\mathrm{B}} = \left(k_x^2 - k_\mathrm{W}^2 - \alpha k_{y}^2 + \alpha \partial_z ^2 + M(z) \right)\sigma_x 
 		+  k_x k_y \sigma_y -i k_x \partial_z \sigma_z,
\end{equation}
where $M(z) = 0$ for $z<0$ and $M(z) = M \rightarrow \infty$ for $z>0$. Here we simplify the model by dropping some constants which will be restored later. 

For the localized surface states, we take the ansatz:
\begin{equation}
\psi (z)\propto \left\{ 
\begin{array}{cc}
\left( 
\begin{array}{c}
u \\ 
iv%
\end{array}%
\right) e^{\lambda _{<}z}, & z<0 \\ 
\left( 
\begin{array}{c}
u \\ 
iv%
\end{array}%
\right) e^{-\lambda_{>}z}, & z>0%
\end{array}%
\right. ,
\end{equation}
where $\mathrm{Re} \lambda _{<}$ and $\mathrm{Re} \lambda _{>}$ are positive. For $z>0$, in the limit of $M \rightarrow \infty$, we have, by taking the ansatz into $H_{\mathrm{B}}\psi = 0$ and neglecting small numbers, 
\begin{equation}
\left( M + \alpha \lambda_{>}^2 \right) u -k_x \lambda_{>} v =0 ,
\end{equation}
\begin{eqnarray}
\left( M + \alpha \lambda_{>}^2 \right) u -k_x \lambda_{>} v  =0 , \\
\left( M + \alpha \lambda_{>}^2 \right) v -k_x \lambda_{>} u =0 ,
 		\label{model_H_B_surf}      
\end{eqnarray}
leading to 
\begin{equation}
k_x \lambda_{>} v  = \pm \left( M + \alpha \lambda_{>}^2 \right) .
\end{equation}
Since $\lambda_{>}>0$, we have $u=v$ for $k_x>0$ and $u=-v$ for $k_x<0$. As a result, we have boundaries conditions as
\begin{equation}
\psi (z=0)\propto \left( \begin{array}{c} 1 \\ \pm i \end{array} \right) \mathrm{for ~} k_x>0~ (k_x<0).  
\end{equation}
With these boundary conditions, we take the ansatz for $z<0$ into $H_{\mathrm{B}}\psi = E\psi$ and we have, for $k_x>0$,
\begin{equation}
i\left(k_x^2 - k_\mathrm{W}^2 - \alpha k_{y}^2 + \alpha  \lambda_{<} ^2 \right) + k_x k_y +  i\lambda_{<} k_x =  E.
\end{equation} 
Equating the real parts and the imaginary parts separately, we have
$E=k_x k_y$ and 
\begin{equation}
\lambda_{<} = \frac{1}{2\alpha} \left\lbrace -k_x  + \sqrt{k_x^2 - 4 \alpha \left(k_x^2 -k_\mathrm{W}^2 - \alpha k_y^2 \right)}\right\rbrace. 
\end{equation} 
In order to have  $\mathrm{Re} \lambda_{<}>0$, $k_x$ is limited by
$k_x < \sqrt{k_\mathrm{W}^2 +  \alpha k_y^2 }$. Similarly, for $k_x<0$, we have $E=-k_x k_y$ when $ -\sqrt{k_\mathrm{W}^2 +  \alpha k_y^2 }< k_x$.
In conclusion, when putting back omitted constants, the surface states (the Fermi arcs) survive in $|k_x| <\sqrt{k_\mathrm{W}^2 +  \alpha k_y^2 } $ and take energy 
\begin{equation}
E = \frac{v_{\parallel}}{k_\mathrm{W}} |k_x| k_y.
\end{equation}
The corresponding wave functions are
\begin{equation}
\psi (z<0) = \lambda_{\mathrm{B}} e^{i k_x x} e^{i k_y y} e^{\lambda_{\mathrm{B}} z} \frac{1}{\sqrt{2}}\left( \begin{array}{c} 1 \\ \mathrm{sgn}(k_x) i \end{array} \right), 
\end{equation}
where 
\begin{equation}
\lambda_{\mathrm{B}} = \frac{m}{\alpha} \left\lbrace - \frac{v_{\parallel}}{k_\mathrm{W}} |k_x|  + \sqrt{ \left(\frac{v_{\parallel}}{k_\mathrm{W}}k_x \right)^2 - \frac{\alpha}{m^2} \left(k_x^2 -k_\mathrm{W}^2 - \alpha k_y^2 \right)}\right\rbrace. 
\end{equation} 

\subsection{Model A}
With the same trick one can also solve the surface states for Model A. We skip the deviations and only show the result below:
for $|k_x|<\sqrt{k_\mathrm{W}^2 +  \alpha k_y^2 }$
\begin{equation}
\psi (z<0) = \lambda_{\mathrm{A}} e^{i k_x x} e^{i k_y y} e^{\lambda_{\mathrm{A}} z} \frac{1}{\sqrt{2}}\left( \begin{array}{c} 1 \\ -i \end{array} \right), 
\end{equation}
where
\begin{equation}
\lambda_{\mathrm{A}} = \frac{m}{\alpha} \left\lbrace - v_{\parallel}  + \sqrt{ v_{\parallel}^2 - \frac{\alpha}{m^2} \left(k_x^2 -k_\mathrm{W}^2 - \alpha k_y^2 \right)}\right\rbrace. 
\end{equation} 
with energy $E= - v_{\parallel} k_y$.

We point out main difference between the two models. Model A shows typical understanding of a Fermi arc connecting two Weyl nodes with linear dispersion $E \propto k_y$. However, in Model B the surface-state wave function is not continuous at $k_x=0$, which is proportional to $\left(1 ,i \right)^{T}$ on one side and $\left(1 ,-i \right)^{T}$ on the other. So it indicates that there are two Fermi arcs that do not connect these Weyl nodes (but to other pairs), presenting a hyperbola dispersion $E \propto |k_x| k_y$.

% ----------------------------- %
% 		Weyl node annihilation		%
% ----------------------------- %
\section{Annihilation of Weyl nodes}
The pair of Weyl nodes move toward each other if we tune the parameter $k_\mathrm{W}^2$ smaller and eventually will collide with each other and annihilate. Before collision, the two Weyl nodes remain intact and the system is still gapless for both Model A and Model B. However, the outcomes are different for the two models after the collision, i.e. $k_\mathrm{W}^2 < 0$. 
For Model A, the system will be gapped out with minimum energy formed by a ring. On the contrary, model B still remain gapless, while the two nodes annihilate into a nodal ring with zero energies. To demonstrate this more clearly, we rescale the parameters to simplify the Hamiltonian for the two models as
\begin{equation}
 H_\mathrm{A} = [k_x^2 -  k_\mathrm{W}^2 - \alpha(k_y^2 + k_z^2) ] \sigma_x
              + k_y \sigma_y +  k_z \sigma_z     
\end{equation}
\begin{equation}
 H_\mathrm{B} = [k_x^2 -  k_\mathrm{W}^2 - \alpha(k_y^2 + k_z^2) ] \sigma_x
              + k_x k_y \sigma_y + k_x k_z \sigma_z   
\end{equation}
Before collision where $k_\mathrm{W}  > 0$, the Weyl nodes are located at 
$(\pm k_\mathrm{W}, 0, 0)$. By tuning $k_\mathrm{W}^2$ to become negative, 
Model A has gap $E_g = 2|k_\mathrm{W}|$ at $k=(0, 0, 0)$ or gap of $E_g= \frac{\sqrt{ 4\alpha |k_\mathrm{W}^2| - 1}}{\alpha}$ at ring of $\sqrt{k_y^2 + k_z^2} =\sqrt{\frac{2\alpha k_\mathrm{W}^2 -1 }{2\alpha^2}}$ 
for $k_x=0$ and other positions are also gapped. Model A is fully gapped unless accidental cases like $|k_\mathrm{W}^2| = \frac{1}{4\alpha}$.  
However, Model B remains gapless in which the WNs annihilate into a nodal ring which has zero energies. 
The nodal ring has the radius $k=\sqrt{\frac{|k_\mathrm{W}^2|}{\alpha}}$ in the $k_x=0$ mirror plane.

\subsection{The introduction of mirror symmetry breaking effect}
We can further see the effect of mirror symmetry breaking terms induced by some perturbations to the Weyl nodes before and after the collision. Such perturbation can be realized several way, such as applying magnetic field to the system. Depending on specific material system Hamiltonian, the spin operators can in different combination of Pauli matrices, determined by system symmetries. Since the mirror operators $\mathcal{M}_x$ for Model A and Model B are known, in conjunction with combined symmetry $\mathcal{C}_{2}\mathcal{T}$, the allowed forms of spin operators can be determined.
Further restriction of allowed forms is possible if we have more symmetry constraints, but this definitely depends on details of the systems. The procedure to determine forms of spin operators that can be concordant with symmetry requirement is shown in the next section.

Here we demonstrate effects of some possible mirror symmetry breaking terms induced by applying magnetic field to the system. In Model A, the allowed spin operator $y$ component can be combination of 
$s_y =\{ k_x \sigma_0, k_x \sigma_x, k_x \sigma_y\}$, which can break mirror symmetry if we apply magnetic field in $y$ direction. For simplicity, we restrict the discussion to the form of $k_x \sigma_y$. The mirror symmetry breaking perturbation is therefore $\Delta k_x \sigma_y$ where the $\Delta$ is small perturbation determined by the strength of magnetic field.  As usual in the 
$k_\mathrm{W}^2 > 0$, the locations of WNs are determined by each component of Pauli matrices to be zeros. The Weyl nodes are still topologically protected to exist but shifted to position of $(\pm \sqrt{\frac{ k_\mathrm{W}^2}{1-\alpha \Delta^2}}, \mp \Delta \sqrt{\frac{ k_\mathrm{W}^2}{1-\alpha \Delta^2}}, 0)$. As for the annihilation results by tuning $k_\mathrm{W}^2 <0$ to make Weyl nodes to collide, the system is still fully gapped where the gap $E_g = 2|k_\mathrm{W}|$ at the origin and $E_g= \frac{\sqrt{ 4\alpha |k_\mathrm{W}^2| - 1}}{\alpha}$ at $k=(0, 0, \pm 
\sqrt{\frac{2\alpha k_\mathrm{W}^2 -1 }{2\alpha^2}})$.

For Model B, the mirror symmetry breaking perturbation can be $\Delta \sigma_y$ induced by the magnetic field if the spin y component $s_y=\sigma_y$ fulfilling the symmetry requirement. The WNs are shifted to $k = (\pm \sqrt{ \frac{k_\mathrm{W}^2 + \sqrt{k_\mathrm{W}^4 + 4\alpha \Delta^2} }{2} }, \mp \sqrt{ \frac{-k_\mathrm{W}^2 + \sqrt{k_\mathrm{W}^4 + 4\alpha \Delta^2} }{2\alpha} }, 0)$, where the relative sign of $k_x$ and $k_y$ are determined by the sign of $\Delta$. After WNs annihilation when tuning $k_\mathrm{W}^2 <0$, the system remains gapless but no longer existing a nodal ring. The zero gapless positions are located at 
$k = (\pm \sqrt{ \frac{ -|k_\mathrm{W}^2| + \sqrt{k_\mathrm{W}^4 + 4\alpha \Delta^2} }{2} }, \mp \sqrt{ \frac{ |k_\mathrm{W}^2| + \sqrt{k_\mathrm{W}^4 + 4\alpha \Delta^2} }{2\alpha} }, 0)$.

Here we demonstrate the differences between Model A and Model B in their behaviour of WNs annihilation when the parameter $k_\mathrm{W}^2$ are tuned from positive to negative. Although both of them describe the pair of WNs when $k_\mathrm{W}^2 >0$, Model A generally will be gapped out when $k_\mathrm{W}^2 <0$ while Model B remains gapless  located at a nodal ring. Even if we apply the magnetic field to break mirror symmetry, the feature of gapful Model A and gapless Model B remains the same. 

% ----------------------------- %
%    spin operator definitions		%
% ----------------------------- %
\section{Allowed forms of spin operators \label{Sec:spin}}
First of all, we have to point out that the Pauli matrices in the Hamiltonians $H_{\mathrm{A}}$ and $H_{\mathrm{B}}$ stand for the pseudo-spin describing two bands' degrees of freedom not real spin. With spin-orbit interaction, spin, orbital and momentum are strongly coupled in bands, so that there is no simple or universal relation between the pseudo-spin and spin. Here we will try to extract spin degrees of freedom from the pseudo-spin based on symmetry point of view and give readers an idea how spin is included in the pseudo-spin for our models. The relation will no doubt depend much on details of systems. 

We start with Model B, in which the mirror operator chosen is $\mathcal{M}_{x}=\sigma_x$. Under the mirror reflection, momentum and spin change as $\mathbf{k}=(k_{x},~k_{y},~k_{z})\mapsto (-k_{x},~k_{y},~k_{z})$ and $\mathbf{s}=(s_{x},~s_{y},~s_{z})\mapsto (s_{x},~-s_{y},~-s_{z})$. At the same time, the pseudo-spin changes as $(\sigma_{x},~\sigma_{y},~\sigma_{z})\mapsto (\sigma_{x},~-\sigma_{y},~-\sigma_{z})$. We find that the pseudo-spin and spin have the same transformation and might conclude that they are identical. However, we cannot have such conclusion because we have not compare their transformations under all possible symmetry operations. As a results, with only the mirror symmetry, we can claim that the spin components might contain contributions as follows:
\begin{align}  
\begin{split}
s_{x} &=\left\lbrace \sigma_0,~k_y \sigma_0,~k_z \sigma_0,~\sigma_x,~k_y \sigma_x,~k_z \sigma_x,~k_x \sigma_y,~k_x \sigma_z \right\rbrace, \\
s_{y} &=\left\lbrace k_x \sigma_0,~k_x \sigma_x,~\sigma_y,~k_y \sigma_y,~k_z \sigma_y,~\sigma_z, ~k_y \sigma_z,~k_z \sigma_z\right\rbrace, \\
s_{z} &=\left\lbrace k_x \sigma_0,~k_x \sigma_x,~\sigma_y,~k_y \sigma_y,~k_z \sigma_y,~\sigma_z, ~k_y \sigma_z,~k_z \sigma_z\right\rbrace, 
\end{split}
\end{align}
where linear combinations of elements in the curly brackets are possible with proper normalization.

To reduce the complexity, we consider that there also exists the combined symmetry of twofold rotation about $z$ and time-reversal symmetry, denoted by $\mathcal{M}_{2}T$. Suppose that the $k_z$ is relative to $k_z=0$ or $\pi$, $\mathcal{C}_{2}\mathcal{T}$ makes $(k_{x},~k_{y},~k_{z})\mapsto (k_{x},~k_{y},~-k_{z})$, $(s_{x},~s_{y},~s_{z})\mapsto (s_{x},~s_{y},~-s_{z})$. Many antiunitary operators could be used for $\mathcal{C}_{2}\mathcal{T}$ with the restriction that \begin{equation}
 \left(\mathcal{C}_{2}\mathcal{T} \right)^2=1,
\end{equation}
either for spin-0 or spin-1/2 systems.
However, when we refer to the transformation
\begin{equation}
\mathcal{C}_{2}\mathcal{T} H_{\mathrm{B}}(k_x,k_y,k_z) \left(\mathcal{C}_{2}\mathcal{T} \right)^{-1} = H_{\mathrm{B}}(k_x,k_y,-k_z),
\end{equation}
we find that it has to be $\mathcal{C}_{2}\mathcal{T} = \sigma_x K$, where $K$ is the complex conjugation operation.
With $\mathcal{C}_{2}\mathcal{T}$, the spin will reduce its compositions as follows:
\begin{align}  
\begin{split}
s_{x} &=\left\lbrace \sigma_0,~k_y \sigma_0,~\sigma_x,~k_y \sigma_x,~k_x \sigma_y \right\rbrace, \\
s_{y} &=\left\lbrace k_x \sigma_0,~k_x \sigma_x,~\sigma_y,~k_y \sigma_y,~k_z \sigma_z \right\rbrace, \\
s_{z} &= \left\lbrace k_z \sigma_y,~\sigma_z, ~k_y \sigma_z \right\rbrace.
\end{split}
\end{align}

With the spirit, we can obtain spin from the pseudo-spin for Model A too. Here $\mathcal{M}_{x}=\sigma_0$ and $\mathcal{C}_{2}\mathcal{T} = \sigma_x K$, we show them as
\begin{align}  
\begin{split}
s_{x} &=\left\lbrace \sigma_0,~k_y \sigma_0,~\sigma_x,~k_y \sigma_x,~\sigma_y, ~k_y \sigma_y \right\rbrace, \\
s_{y} &=\left\lbrace k_x \sigma_0,~k_x \sigma_x,~k_x \sigma_y \right\rbrace, \\
s_{z} &= \left\lbrace k_x \sigma_z \right\rbrace. 
\end{split}
\end{align}

%\end{document}

\end{document}